%% file: main.tex
\documentclass[
prb,
reprint,
twocolumn,
aps,
superscriptaddress,
article
]{revtex4-2}
\usepackage[utf8]{inputenc}
\usepackage{amsfonts}
\usepackage{amsmath}
\usepackage{amsthm}
\usepackage{braket}
\usepackage{graphicx}
\usepackage{hyperref}
\hypersetup{
    colorlinks=true,
    linkcolor=blue,
    filecolor=cyan,      
    urlcolor=magenta,
    citecolor=purple,
}
\usepackage[dvipsnames]{xcolor}
\usepackage{amssymb}
\usepackage{dsfont}
\usepackage{makecell}
\usepackage{verbatim}
\usepackage{amsmath,amscd}
\usepackage{multirow}
\usepackage[normalem]{ulem}
\usepackage{bm}
\usepackage{tikz-feynman}
\usepackage[caption=false]{subfig}
\def \k{{\mathbf k}}
\def \p{{\mathbf p}}
\def \q{{\mathbf q}}
\def \Q{{\mathbf Q}}

\def \H{{\mathbf H}}

\def \r{{\mathbf r}}

\def \v{{\mathbf v}}

\def \S{{\mathbf S}}

\def \beq{\begin{eqnarray}}
\def \eeq{\end{eqnarray}}
\def \nn{\nonumber \\}

\newcommand{\supplementarysection}{%
  \setcounter{figure}{0}
  \let\oldthefigure\thefigure
  \renewcommand{\thefigure}{S\oldthefigure}
  \setcounter{section}{0}
  \let\oldthesection\thesection
  \renewcommand{\thesection}{\oldthesection}
  \setcounter{equation}{0}
  \let\oldtheequation\theequation
  \renewcommand{\theequation}{S\oldtheequation}
  \setcounter{table}{0}
  \let\oldthetable\thetable
  \renewcommand{\thetable}{S\oldthetable}
}

\begin{document}

\title{Field-induced magnon decays in dipolar quantum magnets}

\author{Andrew D. Kim}
\affiliation{Department of Physics, Carnegie Mellon University, Pittsburgh, PA 15213, USA}

\author{Ahmed Khalifa}
\affiliation{Department of Physics, Carnegie Mellon University, Pittsburgh, PA 15213, USA}

\author{Shubhayu Chatterjee}
\affiliation{Department of Physics, Carnegie Mellon University, Pittsburgh, PA 15213, USA}

\begin{abstract}
We investigate the spontaneous disintegration of magnons in two-dimensional ferromagnets and antiferromagnets dominated by long-range dipolar interactions. 
Analyzing kinematic constraints, we show that the unusual dispersion of dipolar ferromagnets in a uniform magnetic field precludes magnon-decay at all fields, in sharp contrast to short-range exchange-driven magnets.
However, in a staggered magnetic field, magnons can decay in both dipolar ferromagnets and antiferromagnets.
Remarkably, such decays do not require a minimum threshold field, and happen over a nearly fixed fraction of the Brillouin Zone in the XY limit, highlighting the significant role played by dipolar interactions.  
In addition, topological transitions in the decay surfaces lead to singularities in the magnon spectrum. 
Regularizing such singular behavior via a self-consistent approach, we make predictions for dynamical spin correlations accessible to near-term quantum simulators and sensors. 
\end{abstract}

\maketitle
The dynamics of a quantum phase of matter are typically governed by its low-energy excitations.
These elementary excitations, such as phonons in solids, electrons or holes in metals, and magnons in ferromagnets or antiferromagnets, are conventionally assumed to be stable and long-lived \cite{AshcroftMermin,simonBook,zimanBook}. 
However, collisions with other quasiparticles or spontaneous disintegration into two or more quasiparticles can render a quasiparticle unstable \cite{lifshitzPitaevskiiBook}. 
At low temperatures, the vanishing density of quasiparticles implies that an excitation cannot decay via collisions. 
Nevertheless, subtle quantum many-body effects can still lead to spontaneous decay of a quasiparticle.
Such decay can have drastic consequences, such as a termination of the spectrum in bosonic superfluids \cite{pitaevskii1959,lifshitzPitaevskiiBook,Graf1974,Glyde1998}, or the destruction of long-lived excitations at the Fermi surface in non-Fermi liquids \cite{schofield1999NFL,senthil2008critical,NFL_Sungsik,NFL_Chowdhury}, that render the quasiparticle desciption invalid. 

In short-range interacting quantum magnets with collinear magnetic order, e.g. XY ferromagnets and N\'eel antiferromagnets, external magnetic fields are known to induce magnon decays \cite{Zhitomirsky1999,Mourigal2010,Stephanovich_2011,ZC_RMP2013}. 
For instance, in the Heisenberg antiferromagnet on a cubic lattice, all magnons remain perturbatively stable up to a threshold magnetic field $H^*$. 
As one increases $H$ beyond $H^*$, an increasingly larger fraction of magnons in the Brillouin Zone (BZ) decay spontaneously, leading to a marked deviation from mean-field predictions for spin dynamics \cite{Zhitomirsky1999,Mourigal2010,ZC_RMP2013}. 
\begin{figure}[!t]
    \centering
    \includegraphics[width=1.0\linewidth]{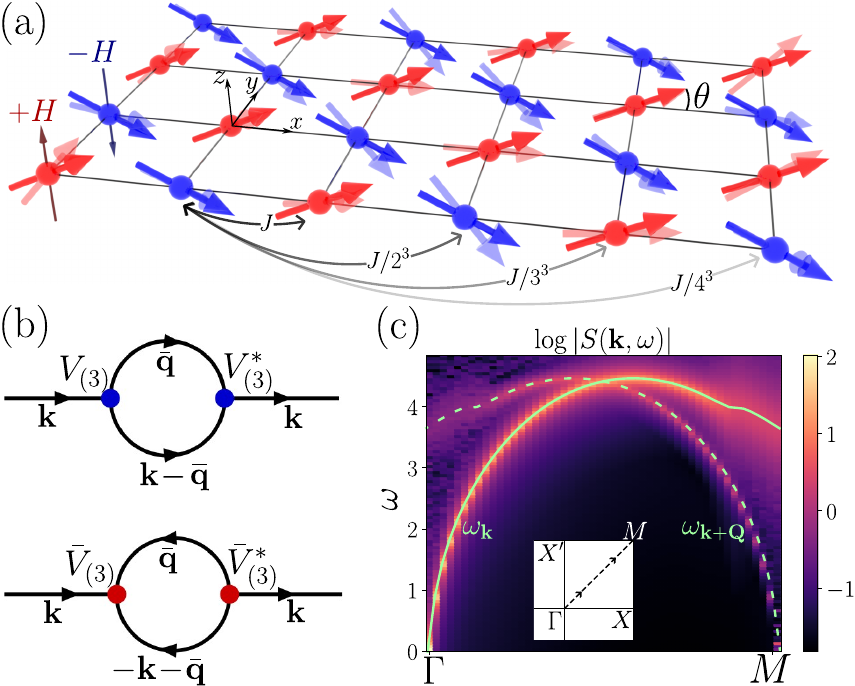}
    \caption{(a) Schematic depiction of the canted ground state of dipolar XXZ ferromagnet in a staggered magnetic field. The lightly colored arrows correspond to a magnon texture. (b) One-loop self-energy diagrams generated by cubic vertices. (c) Log plot of dynamical structure factor as defined in text along a line cut in the BZ (inset), showing mixing of magnon modes at $\k$ (solid line) and $\k + \Q$ (dashed line) due to staggering, and blurring of spectral features at higher energies due to magnon decays.}
    \label{fig1:schematic}
\end{figure}

Recent experiments in both synthetic and material platforms point to regimes where dipolar ($1/r^3$) interactions dominate. 
Many near term quantum simulators, such as Rydberg atom arrays \cite{Li2005,browaeys2020,CSB_Rydberg}, driven trapped ion chains \cite{jurcevicQuasiparticleEngineeringEntanglement2014,richermeNonlocalPropagationCorrelations2014}, and ultracold polar molecules \cite{gadway2016strongly,baoDipolarSpinexchangeEntanglement2022,christakisProbingSiteresolvedCorrelations2023}, naturally realize a resonant dipolar interaction between effective spin-degrees of freedom.
On the materials front, dipolar couplings describe lattices of charge-neutral excitons with perpendicular electrical dipole moments in 2D moir\'e heterostructures \cite{Gu2022dipolar,Yu2023,Lian2023quadrupolar,lian2024valley} and are argued to be important for spin dynamics in certain Van der Waals magnets  \cite{Sun2024DipolarAntiferromagnet,Ziffer24}. 
While a flurry of work \cite{HPB2012,Diessel_2023,Giachetti2022,LRI_RMP2023,Sbierski2023,Dipolar_Longpaper} has investigated the effects of dipolar interactions on the ordered ground state, the dynamics of these quantum dipoles remain less explored \cite{Frerot2017}. 
Specifically, the long-range tail of the dipolar interaction fundamentally alters the dispersion of magnons \cite{HPB2012}, which can affect their stability.
Understanding potential instabilities of magnons is a necessary pre-requisite to decipher the quantum dynamics of spins in long-range interacting magnets.

In this paper, we study the stability of magnons in long-range interacting quantum magnets in the presence of an external magnetic field.
Specifically, we focus on quantum fluctuation-induced decays at zero temperature in 2D collinear ferromagnets (FM) and antiferromagnets (AFM) on the square lattice, and consider both uniform and staggered magnetic fields motivated by  recent experiments in Rydberg arrays \cite{CSB_Rydberg,Chen2023}. 
Our central results are as follows. 
First, using non-linear spin-wave theory, we show that in dipolar XY FMs, magnons are absolutely stable in uniform magnetic fields due to kinematic constraints. 
At non-zero staggered fields, magnons can decay.
Such decays occur at infinitesimal fields, and the decay region does not evolve as a function of the field, both in contrast to previously studied short-range interacting magnets in uniform fields. 
Remarkably, dipolar XY AFMs, expected to behave like short-range interacting magnets as the dipolar interactions frustrate AFM order, also lack a threshold field $H^*$ for magnon decay.
Further, for dipolar XY AFMs in staggered fields, the decay region decreases with increasing $H$, in contrast to short range AFMs.
Second, we identify singularities in the magnon spectrum due to divergent decay-rates. 
We show that such singularities originate from a transfer of spectral weight to the single magnon mode from the two-magnon continuum, and correspond precisely to topological transitions in the decay contour with a Van Hove singularity in the two-magnon density of states.
Third, we investigate the implications of magnon decay, and demonstrate that the dynamical spin correlations show striking deviations from mean-field predictions. 

\emph{Model}.- We consider generic dipolar spin-$S$ XXZ Hamiltonians on a square lattice in a magnetic field.
\beq
\hat{\cal{H}} = \frac{J}{2} \sum_{i \neq j} \frac{1}{r_{ij}^3} \left[ S_i^{x}  S_j^{x} + S_i^{y}  S_j^{y} + \Delta S_i^{z}  S_j^{z} \right] - \sum_{i} H_i S_i^{z} ~~
\label{eq:H}
\eeq
where $\Delta$ is the spin anisotropy that tunes between the XY $(\Delta  = 0)$ and Heisenberg $(\Delta = 1)$ limits, and $H_i$ is an external magnetic field which we take to be either uniform $(H_i = H)$ or staggered ($H_i = e^{i \Q \cdot \r_i} H$, with $\Q = (\pi, \pi)$).
This model features an ordered ground state which spontaneously breaks a continuous spin-rotation symmetry.
At $H = 0$, the ground state is a collinear FM (AFM) when $J < 0$ ($J > 0$) (Fig.~\ref{fig1:schematic}(a)), and features gapless magnons that correpond to Goldstone modes of the spontaneously broken spin-rotation symmetry.
When $H \neq 0$, the spins orient in the $xy$ plane and then cant in the $z$ direction, leading to a non-collinear magnetic order (Fig.~\ref{fig1:schematic}(a)), while still spontaneously breaking a U(1) symmetry generated by $\sum_i S^{z}_i$. 
Beyond a critical saturation field $H = H_c$, the spins all align with the uniform (staggered) magnetic field, and a trivial field-induced FM (AFM) is obtained, with gapped magnons.

To study the spectrum and stability of magnons, we resort to non-linear spin-wave theory. 
To this end, we follow the standard procedure \cite{ZC_RMP2013} of rewriting $\hat{\cal{H}}$ in the local rotating reference frame corresponding to the classical ground state, such that each spin is aligned along the local $x$ axis \cite{supp}.
Next, we rewrite the spin $S^\alpha_j$ operators in terms of Holstein-Primakoff bosons $a_j$ as 
\begin{equation}
S^x_j = S - a^\dagger_j a_j, S^+_j = (2S - a^\dagger_j a_j)^{1/2}a_j, S^-_j = (S_j^+)^\dagger
\end{equation}
Subsequently, a canonical Bogoliubov transformation to bosons $b_\k$ yields the magnon dispersion $\omega_\k$ at quadratic order, as well as interactions between magnons in a 1/S expansion.
\beq
\hat{\cal{H}} &=& \sum_{\k} \omega_\k b^\dagger_\k b_\k + \frac{1}{\sqrt{N}} \sum_{\k, \q} \bigg[ \frac{V^{(3)}_{\k,\q}  }{2!}b^\dagger_\k b_\q b_{\k -\bar{\q}} + 
\frac{\bar{V}^{(3)}_{\k,\q} }{3!} b^\dagger_{\k} b^\dagger_\q b^\dagger_{-\bar{\q} - \k} \nn
&& + \mbox{H.c.} \bigg] + \frac{1}{N} \sum_{\k,\q,\p} 
\frac{ V^{(4)}_{\k,\q,\p}}{3!} b^\dagger_{\k} b_{\p} b_{\q} b_{\k - \p -\q} +  \mbox{H.c.}  + \ldots
\label{eq:Hb}
\eeq

In Eq.~\eqref{eq:Hb}, $V^{(3)}_{\k,\q}$ is the leading-order 3-magnon scattering vertex: it is symmetry forbidden for collinear magnets \footnote{We caution the reader that this conclusion no longer holds if one has bond-dependent anisotropic exchange, \cite{winter2017breakdown,Smit2020,Tao2023}}, and hence only appears when field-induced canting makes the magnetic texture non-collinear and mixes longitudinal and transverse modes \cite{Zhitomirsky1999,Mourigal2010,ZC_RMP2013,Stephanovich_2011,CZ_triangularPRL,CZ_triangular2009}. 
Further, $\bar{\q} = \q$ for the FM in a uniform field, otherwise $\bar{\q} = \q - \Q$, where $\Q$ is related to staggering due to the external field (FM) or simply the spontaneous magnetic order (AFM) --- this plays an important role in determining the kinematic constraints on magnon decay \footnote{We always work in the full BZ of the square lattice, so $\q$ and $\bar{\q} = \q - \Q$ are distinct momenta even in the presence of a staggered field.}. 
$V^{(4)}_{\k,\p,\q}$ is a 4-magnon vertex that does not involve any staggering factor and is generically present in both collinear and non-collinear magnets, although its magnitude is suppressed by $1/\sqrt{S}$ relative to $V^{(3)}_{\k,\q}$. 
Finally, the ellipsis denotes the classical energy, 4-boson terms that do not contribute to spontaneous magnon decay at $T =0$, and higher order terms in 1/S \cite{supp}.

Hamiltonian in hand, we can now use tools of diagrammatic perturbation theory to evaluate the stability of magnons in dipolar quantum magnets \cite{ZC_RMP2013}. 
To this end, we define the magnon Green's function $G(\k,t) = - i \langle T(b_\k(t) b_\k^\dagger(0)) \rangle$. 
Following a Fourier transform, the mean-field Green's function is simply given by the mean-field magnon dispersion: $G^{(0)}(\k,\omega) = (\omega - \omega_\k)^{-1}$. 
To compute leading quantum corrections to the magnon dispersion, we solve the Dyson equation for $G(\k,\omega)$ \cite{Coleman_2015} by evaluating the one loop self-energy $\Sigma(\k,\omega)$ generated by the 3-boson vertex (Fig.~\ref{fig1:schematic}(b)) \footnote{Here, we neglect anomalous contributions arising from vertex corrections. We also neglect Hartree-Fock like corrections from the four-magnon terms; for justification, see Section \ref{supp_section_quartic_terms_3magnon_decays} of the Supplemental Material}.
The renormalized magnon Green's function is given by $G(\k,\omega) = \left(\omega - \omega_\k - \Sigma(\k,\omega)\right)^{-1}$.
The real part of the on-shell self-energy $\Sigma(\k, \omega_\k)$ shifts the magnon spectrum, while its imaginary part, which we will mainly focus on, leads to a finite lifetime for the magnons. 
In particular, a large imaginary component of $\Sigma(\k, \omega_\k)$ corresponds to fast decay of the magnon at momentum $\k$, and a divergence indicates that magnons are unstable. 
Two key factors determine the self-energy induced by 
3-boson terms in $\cal{\hat{H}}$ --- (i) the vertex function $V^{(3)}_{\k, \q}$, and (ii) kinematic constraints on energy and momentum conservation. 
Since the vertex is generically non-zero in the BZ, we next turn our attention to kinematic constraints, which furnish a wealth of useful information about magnon decays. 

\emph{Kinematic constraints}.-
The distinct dispersion of magnons in dipolar quantum FMs ($\omega_\k \sim \sqrt{k}$ for $\Delta < 1$)  indicates that the kinematic constraints on magnon decay are substantially different, even without any magnetic field. 
Therefore, we begin with an analysis of the constraints imposed by fundamental conservation laws when $H = 0$.
In this case, a combination of time-reversal and odd parity under $\pi$-rotation about the in-plane axis orthogonal to the ordering direction \footnote{Let the spins be ordered along $\hat{x}$, then $\S \xrightarrow{\Theta} -\S \xrightarrow{R_{\pi}(\hat{y})} (S^x, -S^y, S^z) \implies S^{\pm}= S^y \pm i S^z \xrightarrow{\Theta R_\pi(\hat{y})} -S^{\pm}$} implies that the cubic vertex $V^{(3)}_{\k,\q}$ vanishes. 
Nevertheless, it is useful to analyze the kinematic constraint for a single magnon decaying into two magnons. Stability of magnons against two-particle decays over the entire BZ implies that higher order decay processes, such as one magnon decaying into three magnons, are also forbidden by energy conservation, even though the corresponding vertex  $V^{(4)}_{\k,\p,\q}$ may be non-zero. 

\begin{figure}[!t]
   \centering
   \includegraphics[width=1.0\linewidth]{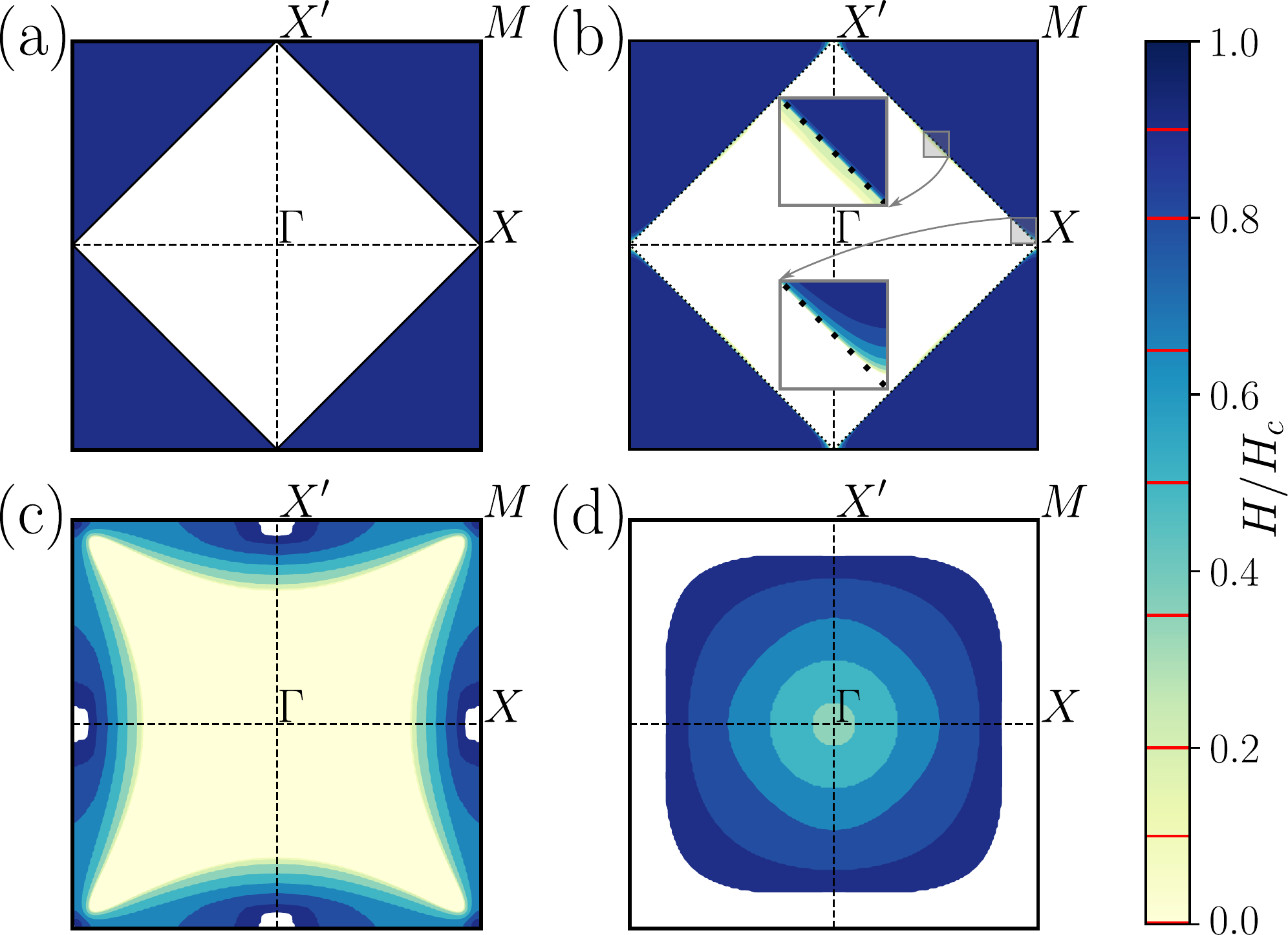}
   \caption{Regions in BZ with kinematically allowed two-magnon decays for different values of $H/H_c$ for (a) staggered-field XY FM, (b) staggered-field XY AFM, (c) uniform-field XY AFM, and (d) uniform-field Heisenberg AFM. Upper and lower insets of (b) show deviation of staggered-field XY AFM decay region boundary from the FM boundary.}
    \label{fig2:decay_regions}
\end{figure}

For dipolar FMs, we numerically find that $\omega_\k = \omega_{\q} + \omega_{\k - \q}$ has no solution for any initial momentum $\k$ (excluding the trivial solution $\q = 0$).
This indicates that magnons in dipolar FMs are absolutely stable quasiparticles, in contrast to short-range XXZ FMs where magnons can indeed decay even at zero external magnetic field \cite{Stephanovich_2011}.
Intuitively, this strong restriction on decay is a consequence of the steep magnon dispersion $\omega_\k \sim \sqrt{k}$ caused by long-range dipolar interactions \cite{HPB2012,EpsteinZeta,supp}, which differs drastically from the generic linear dispersion of short-range interacting XXZ magnets \cite{AuerbachBook}.

While turning on a magnetic field $H$ modifies magnon dispersions \cite{supp}, kinematic constraints continue to forbid magnon-decay for dipolar FMs in uniform fields. 
However, for FMs in presence of a staggered magnetic field, as well as for AFMs in either uniform or staggered fields, the kinematic constraint gets modified to $\omega_\k = \omega_{\q} + \omega_{\k - \q + \Q}$, as exemplified by the three-magnon vertex in Eq.~\eqref{eq:Hb}.
Physically, this implies that such a magnon-decay process involves an absorption of a particle into the condensate with momentum $\Q$ due to the staggered nature of the ground state.

For dipolar FMs, the only allowed decays occur via emission of Goldstone magnons \footnote{We do not call these modes acoustic as they do not disperse linearly.} near $\q = 0$ or $\q = \k + \Q$, such that one of the final magnons has nearly zero energy, irrespective of the strength of the applied field $H$.
Consequently, the decay threshold boundary, defined as the contour in the BZ that separates the stable magnons from the unstable ones, is determined by the condition $\omega_\k = \omega_{\k + \Q}$.
On the square lattice, $C_4$ symmetry of the dispersion $\omega_\k$ constrains this condition to be satisfied on the diamond-shaped contour defined by $|k_x \pm k_y| = \pi$ (Fig.~\ref{fig2:decay_regions}(a)).
Any magnon with initial momentum $\k$ inside this contour is therefore stable against decay via quantum fluctuations for any $H < H_c$.

For dipolar AFMs, one expects behavior similar to short-range interacting magnets, as the dipolar interaction is frustrated and effectively short-ranged. 
Surprisingly, in the XY AFM in a staggered field, we not only find a lack of threshold magnetic field (as in the dipolar ferromagnet), but we also find that the allowed decay region \textit{decreases} slightly with increasing field (Fig.~\ref{fig2:decay_regions}(b)). 
Both of these features stand in contrast to the conventional wisdom for short-range interacting magnets \cite{ZC_RMP2013} and illustrate the crucial role played by dipolar interactions in the dynamics of magnetization at intermediate energy-scales. 
Only in a uniform field near the Heisenberg limit does the decay spectrum begin to bear qualitative similarity to short-range magnets, exhibiting both a threshold field $H^* \approx 0.29 H_c$ for the Heisenberg AFM (HAFM) and increasing magnon instability as $H$ is increased beyond $H^*$.
Our results for these different cases are summarized in Table~\ref{tab}.

\def\chk{\tikz\fill[scale=0.4,fill={rgb:red,2;green,2;blue,2}](0,.35) -- (.25,0) -- (1,.7) -- (.25,.15) -- cycle;} 
\def\x{\tikz\fill[scale=0.4,fill={rgb:red,1;green,1;blue,1}](0,0) -- (.325,.4) -- (0,.8) -- (.4,.475) -- (.8,.8) -- (.475,.4) -- (.8,0) -- (.4,.325) -- cycle;} 

\begin{table}[!t]
    \centering
    \begin{tabular}{|c|c|c||c|c|}\hline
        & \multicolumn{2}{c||}{Staggered $H$} & \multicolumn{2}{c|}{Uniform $H$} \\ \hline
        & XY FM  & XY AFM & XY AFM  & HAFM  \\ \hline
        Fixed decay region & \chk & \x & \x & \x \\ \hline
        Threshold $H^*/H_c$ & 0.00 & 0.00 & 0.00 & $0.29$ \\ \hline
    \end{tabular}
    \caption{Magnon decay properties for dipolar XY and Heisenberg magnets in uniform and staggered fields.}
    \label{tab}
\end{table}

\begin{figure}[!t]
   \centering
   \includegraphics[width=1.0\linewidth]{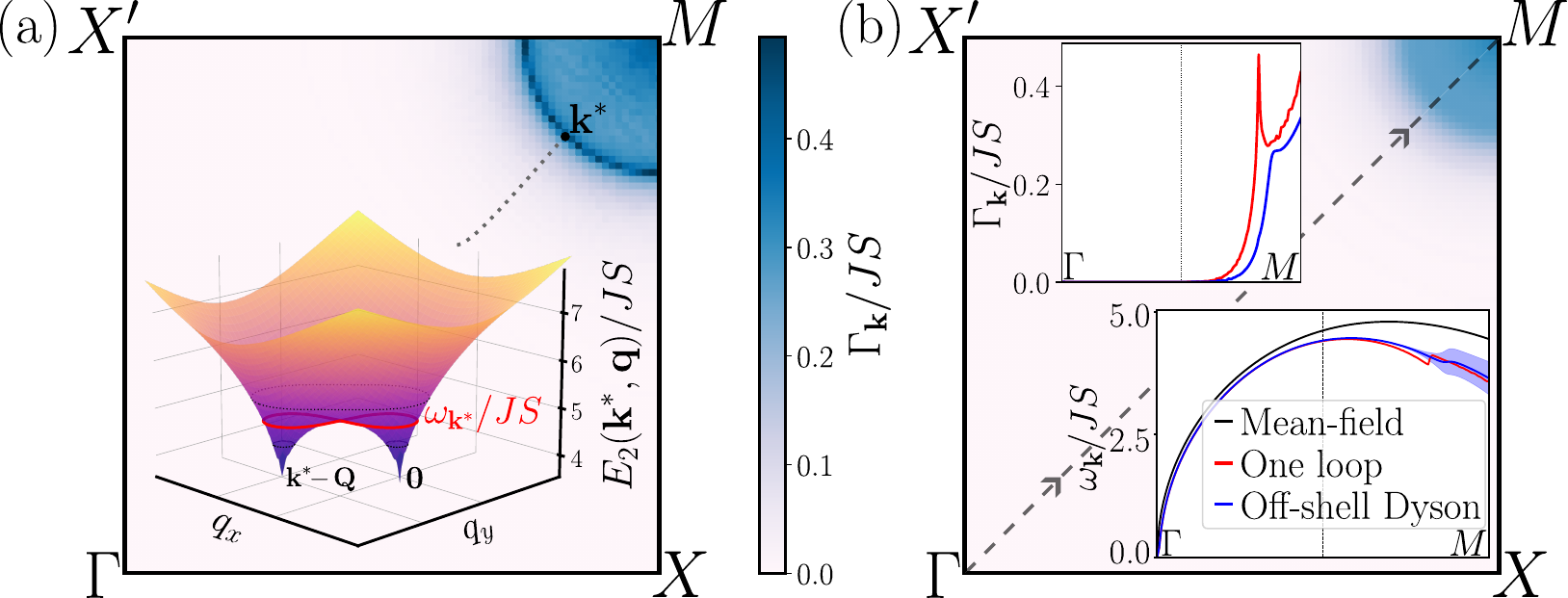}
   \caption{(a) One loop decay rate for the XY FM in a staggered field at $H/H_c = 0.5$ in first quadrant of BZ. Inset: plot of the two-magnon continuum $E_2(\k^*,\q) \equiv \omega_\q + \omega_{\k^*-\q+\Q}$ for a singular $\k^*$, illustrating the topological transition in the decay surface in red. (b) Regularized decay rate from solving off-shell Dyson's equation. Insets: Line cuts showing the decay rate $\Gamma_\k$ (top), and single-magnon dispersions with blue areas indicating width of spectral peaks due to decay (bottom).}
    \label{fig3:stag_XY_FM}
\end{figure}

\emph{Decay rates, threshold behavior and singularities}.-
The effect of spontaneous magnon-decay is encapsulated in the magnon spectral function $A(\k,\omega) = - 2 \, \text{Im}[G(\k,\omega)]$ \cite{Coleman_2015}, which we now turn our attention to.
For the dipolar FM, decays of long-wavelength magnons are forbidden, and the spectral function remains sharply peaked.
Magnons beyond the decay threshold contour, with wavelength comparable to the lattice scale, are allowed to decay. 
The decay rate is given by the imaginary part of the on-shell self-energy: $\Gamma_\k = - \text{Im}[\Sigma(\k,\omega_\k)]$ \cite{Zhitomirsky1999,Mourigal2010,Stephanovich_2011,ZC_RMP2013}.
Consequently, the magnon spectral function takes the form of a Lorentzian centered at $\omega_\k$ with width $\Gamma_\k$. 
Hence, we evaluate $\Gamma_\k$ using a one-loop approximation (Fig.~\ref{fig1:schematic}(b)) and plot it in Fig.~\ref{fig3:stag_XY_FM}(a).

Two features of the decay-rate are apparent from Fig.~\ref{fig3:stag_XY_FM}.
First, we note a strong suppression of the decay rate near the decay threshold.
To quantify this observation, we consider a magnon near the threshold, with initial momentum $\k_0 + \p$ where $\k_0$ lies on the decay boundary, and analytically compute the decay rate in the small $p$ limit.
Since one of the two final magnon states is constrained to have small momentum $\q$ for threshold decays, the decay rate is given by
\beq
\Gamma_{\k_0 + \p} \propto \int d^2q \, |V^{(3)}_{\k_0 + \p,\q}|^2 \; \delta(\p \cdot \hat{\Q} - c \sqrt{q}) \, ,
\label{eq:threshold}
\eeq
where the energy-conserving delta function reflects the kinematic constraint \cite{supp}.
The kinematic constraint sets $q \sim (\p \cdot \hat{\Q})^2$, indicating that the two-magnon density of states is suppressed due to the sharp square-root onset of the single magnon dispersion. 
By contrast, the squared vertex $|V^{(3)}_{\k_0 + \p,\q}|^2 \propto \sqrt{q} \sim |\p \cdot \hat{\Q}|$ vanishes more slowly than short-ranged magnets due to long range dipolar interactions. 
Combining both effects, the decay rate vanishes as $|\p \cdot \hat{\Q}|^{4}$ near the threshold.

Second, we note the presence of a ring centered at the M-point (Fig.~\ref{fig3:stag_XY_FM}(a)) where the decay rate is very large.
This corresponds to a divergent self-energy within the one-loop approximation.
Physically, this coincides with crossing of the single magnon dispersion $\omega_\k$ into the two-particle continuum $E_2(\k,\q)$, and is associated with a topological transition in the decay surface where it fractures into two disjoint pieces, as shown in Fig.~\ref{fig3:stag_XY_FM}(a).
Such a saddle point in $E_2(\k,\q)$ leads to a divergent two-magnon density of states, in turn inducing a divergent decay rate for the single magnon mode via a transfer of spectral weight by the cubic vertex $V^{(3)}_{\k,\q}$.
The corresponding characteristic anomalous behavior may simply be obtained by examining the magnon self-energy in the vicinity of any point at a distance $\Delta k$ away from the ring. 
We find that $\Gamma_\k \approx \ln(\Lambda/\Delta k)$, indicating a logarithmic divergence as $\Delta k \to 0$ ($\Lambda$ is a UV momentum cutoff). 
Concomitantly, this non-analyticity is accompanied by a step-function jump $\Theta(\Delta k)$ in the real part of the self-energy (lower inset of Fig.~\ref{fig3:stag_XY_FM}(b)) that renormalizes the dispersion.

To regularize this divergence, we resort to solving an off-shell Dyson's equation, as discussed in Ref.~\onlinecite{CZ_triangular2009}. 
Intuitively, this amounts to taking into account the fact that the initial magnon is unstable.
Therefore, its spectral function is broadened such that it is able to sample adjacent magnon states in the BZ with non-divergent decay rates, and this regulates the divergence.
Formally, this corresponds to finding a self-consistent solution to the following set of equations:
\begin{align}
\bar{\omega}_\k &= \omega_\k + \text{Re}[\Sigma(\k, \bar{\omega}_\k + i \Gamma_\k)], \nonumber \\
\Gamma_\k &= - \text{Im}[\Sigma(\k, \bar{\omega}_\k + i \Gamma_\k)]
\label{eq:offShellDyson}
\end{align}
The results of solving Eq.~\eqref{eq:offShellDyson} are shown in Fig.~\ref{fig3:stag_XY_FM}(b).
We note that even though this approach regulates the divergence, the magnon-decay rates remain significantly enhanced near the M point, indicating that these magnons no longer remain well-defined quasiparticles.

\begin{figure}[!t]
   \centering
   \includegraphics[width=1.0\linewidth]{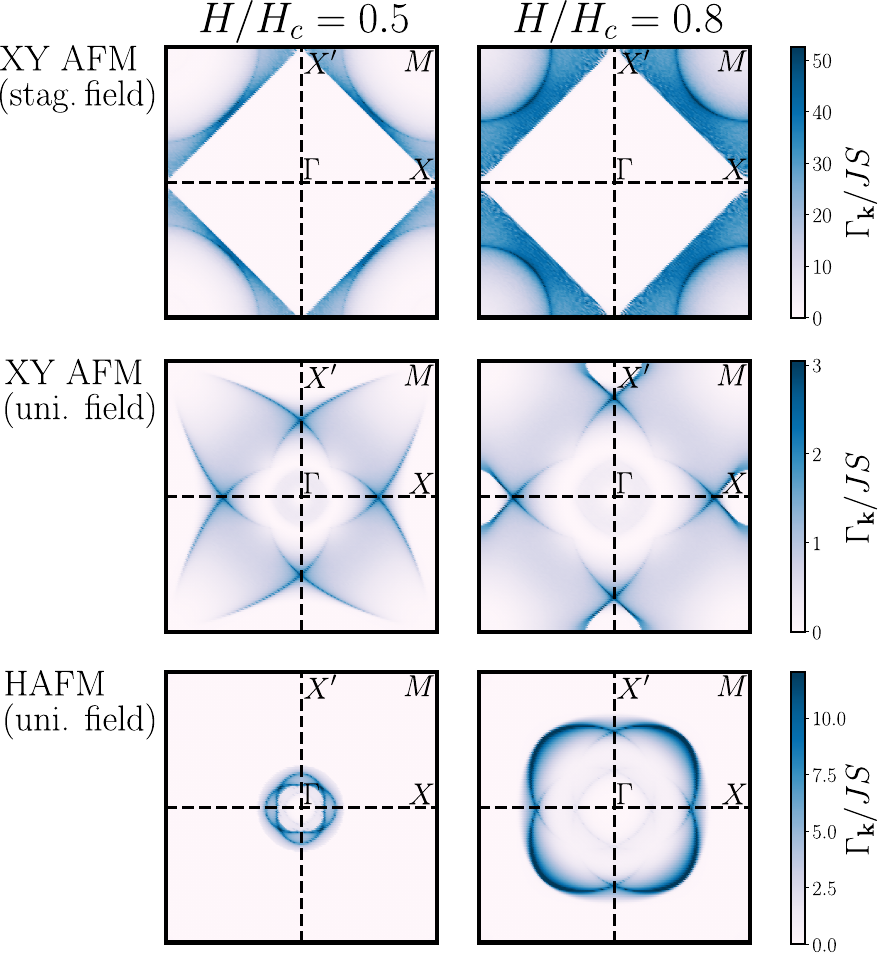}
   \caption{One-loop decay rates for different dipolar AFMs in uniform and staggered fields.}
    \label{fig4:AFM_decays}
\end{figure}

To conclude this section, we contrast our findings with magnon-decay rates for dipolar AFMs in XY and Heisenberg limits for uniform and staggered fields in Fig.~\ref{fig4:AFM_decays}. 
We find that magnon-decays generically happen over a larger fraction of the BZ at high fields, and the decay rates are significantly enhanced relative to the dipolar FM in kinematically allowed regions.

\emph{Dynamical structure factor}.-
Finally, we investigate the physical consequences of magnon-decay by evaluating the dynamical spin-correlations or spin-structure factor $S^{\alpha \beta}(\k, \omega) = \int d^2r \int dt \, e^{i (\k \cdot \r - \omega t)} \langle S^\alpha(\r,t) S^\beta(0,0) \rangle$.
Such a quantity can be readily measured by space-time dependent snapshots of spins for synthetic magnets in arrays of Rydberg atoms or polar molecules \cite{browaeys2020,Ruttley2024,Picard2024}, or by nanoscale magnetometry via color defects in 2D materials \cite{Ziffer24,Xue24,Flebus,CRD19,Nieva22}.
To study the qualitative consequences of the field, we show $S(\k,\omega)= \sum_\alpha |S^{\alpha \alpha}(\k,\omega)|$ in Fig.~\ref{fig1:schematic}(c) for the dipolar XY FM in a staggered field.
We note that the staggered field leads to a redistribution of the spectral weight by mixing the one-magnon (transverse) spectrum at $\k$ and $\k + \Q$.
Further, in the vicinity of the Van Hove singularity, the structure factor broadens out, carrying clear signatures of magnon-instability due to mixing of transverse and longitudinal (two-magnon) modes.
Finally, away from such hybridization, the structure factor is dominated by the sharp single-magnon spectrum, and the contribution of longitudinal modes remains weak. 

\emph{Influence of lattice geometry}.- Our formalism can be directly used to study magnon-dynamics in non-collinear dipolar magnets on non-bipartite lattices \cite{goetze2016ground,Homeier,Liu}, provided magnetic order is not precluded by valence bond solids or quantum spin liquids \cite{Marcus_DSL,Machado_CSL}.
To this end, we investigated magnon-decays in triangular and honeycomb lattices \cite{supp}, and found that dipolar XY ferromagnets in uniform fields are stable in both cases due to severe kinematic constraints. 
However, spontaneous magnon-decays happen for both dipolar ferromagnets and antiferromagnets in a staggered field on the honeycomb lattice, and there is no threshold field $H^*$ just like the square lattice. 
Finally, analogous van Hove singularities in the decay contour lead to significant renormalization of the single-magnon spectrum.  
Our results confirm that several anomalous features of magnon-decay specific to long-range dipolar interactions are broadly applicable across lattice geometries. 

\emph{Conclusion and outlook}.- 
In this work, we critically examined the stability of magnons in dipolar ferromagnets and antiferromagnets on a square lattice in presence of an external magnetic field.  
Our results can be explored both in 2D VdW materials, as well as existing and near-term quantum simulators.
In fact, the stability of ferromagnetic magnons relative to antiferromagnetic magnons at zero-field has been observed in a recent quench experiment in a 2D Rydberg array realizing XY-magnetism \cite{Chen2023}.
Our work predicts that the antiferromagnetic magnons can be stabilized in the Heisenberg limit in a weak uniform field \footnote{Note that this also requires the 1-3 magnon decays to be suppressed by the field, which we demonstrate in the supplement \cite{supp}}, enabling a better experimental characterization of their dispersion.
Our results can also be leveraged to identify appropriate phase spaces for stable magnons, a pre-requisite to transmit information via spin degrees of freedom.

On the theoretical front, our work sets the stage for further investigations of spin-dynamics in long-range interacting magnets. 
While our results apply to the thermodynamic limit, a natural next step is to consider the interplay of magnons with the Anderson tower of states \cite{Tomasso2023} in finite-sized systems of experimental relevance.  
Further, our work opens the door to studying the effect of tunable power-law interactions on magnon dynamics, as realized in driven trapped ion-chains \cite{richermeNonlocalPropagationCorrelations2014}.

\textit{Acknowledgements}. It is a pleasure to acknowledge Marcus Bintz, Antoine Browaeys, Francisco Machado, Lode Pollet, Tomasso Roscilde and Norman Y. Yao for helpful discussions. 
A. D. K. acknowledges support from the Pennsylvania Space Grant Consortium.
A. K. and S. C. acknowledge support from a PQI Community Collaboration Award. 


\bibliography{MagnonDecayLRI}

\newpage

\onecolumngrid

\vspace{0.3cm}

\supplementarysection
 
\input{supp.tex}

\vfill

\end{document}

%% file: supp.tex
\def \sx{S^x}
\def \sy{S^y}
\def \sz{S^z}

\def \eps{\varepsilon}
\def \ac{a^\dag}
\def \aa{a}
\def \bc{b^\dag}
\def \ba{b}
\def \H{\hat{\cal{H}}}

\def \sp{\hspace{0.10in}}

\begin{center}
    \textbf{Supplemental Material: Field-induced magnon decays in dipolar quantum magnets}
\end{center}

\author{Andrew D. Kim}
\affiliation{Department of Physics, Carnegie Mellon University, Pittsburgh, PA 15213, USA}

\author{Ahmed Khalifa}
\affiliation{Department of Physics, Carnegie Mellon University, Pittsburgh, PA 15213, USA}

\author{Shubhayu Chatterjee}
\affiliation{Department of Physics, Carnegie Mellon University, Pittsburgh, PA 15213, USA}

\section{Model Hamiltonians and spin-wave expansion}
\label{supp_section_I}
In this section, we provide additional technical details of non-linear spin-wave theory that is extensively used to study zero-temperature physics of dipolar XXZ Hamiltonians in the main text.
Specifically, we consider dipolar spin-$S$ XXZ Hamiltonians on a square lattice in a magnetic field, which take the form of
\beq
\H = \frac{J}{2} \sum_{i \neq j} \frac{1}{r_{ij}^3} \left[ S_i^{x_0}  S_j^{x_0} + S_i^{y_0}  S_j^{y_0} + \Delta S_i^{z_0}  S_j^{z_0} \right] - \sum_{i} H_i S_i^{z_0} \label{eq:H_appendix} ~~
\eeq
with definitions of $\Delta$ and $H_i$ given in the main text.
One notational difference from the main text is that we use subscripted axes $x_0$, $y_0$, and $z_0$ for spin operators in the lab frame to differentiate from those in a locally rotated frame (which depends on the specific configuration being considered). 

The choices of ferromagnetic (FM) versus antiferromagnetic (AFM) coupling and uniform versus staggered external magnetic field provide four distinct configurations of the dipolar XXZ Hamiltonian. Using spin-wave theory to study them is both straightforward and well-studied, so we only consider one configuration in detail here.

For the staggered field dipolar XXZ FM, we take $H_i = \eta_i H$ and $J>0$, manually inserting a negative sign at the front of the first sum in \eqref{eq:H_appendix} to make the ground-state ferromagnetic.
The ground state has all spins aligned in-plane along an arbitrary direction, which we take to be $\hat{x}_0$ for concreteness.
At nonzero $H$, the ordered moment at each site is expected to cant in the $\pm \hat{z}_0$ direction depending on the sign of $H_i$, as seen in Figure \ref{fig1:schematic}(a) in the main text.
Then, to simplify calculations, we move to a sublattice-dependent twisted reference frame for the spins
\begin{align}
S^{x}_j &= S^{x_0}_j \cos\theta_j + S^{z_0}_j \sin\theta_j \nn
S^{y}_j &= S^{y_0}_j \nn
S^{z}_j &= -S^{x_0}_j\sin\theta_j + S^{z_0}_j\cos\theta_j \label{eq:twisted_frame}
\end{align}
where $\theta_j = \eta_j \theta$ and $\eta_i = e^{i \Q \cdot \r_j}$, such that all spins are aligned along $\hat{x}$ in the twisted frame. 
We then use a Holstein-Primakoff transformation in the twisted frame, introducing bosonic operators $\aa_j,\ac_j$ in the large-$S$ limit as
\begin{align}
    \sx_j &= S - \ac_j \aa_j \nn
    S^+_j &= \sqrt{(2S - \ac_j \aa_j)} \aa_j = \sqrt{2S} \sqrt{ \left(1 - \frac{\ac_j \aa_j}{2S}\right)} \aa_j \approx \sqrt{2S} \left( 1 - \frac{\ac_j \aa_j}{4S} + \cdots \right) \aa_j \nn
    S^-_j &= \ac_j \sqrt{(2S - \ac_j \aa_j)} \approx \sqrt{2S} \ac_j \left( 1 - \frac{\ac_j \aa_j}{4S} + \cdots \right) \label{eq:HP_trans}
\end{align}
where the final, approximate equalities in $S_j^{\pm}$ come from assuming that $n_j/2S \ll 1$, where $n_j \equiv \ac_j \aa_j$ is the boson number operator. Spin operators $S^\alpha_j$ in the twisted frame, to leading order in the above expansion, are then
\beq
    \sx_j = S-\ac_j\aa_j, \sy_j \approx \frac{\sqrt{2S}}{2}(\aa_j + \ac_j), \sz_j \approx \frac{\sqrt{2S}}{2i}(\aa_j - \ac_j). \label{eq:spin_ops}
\eeq 

Substitution of twisted-frame spin operators \eqref{eq:spin_ops} into the XXZ Hamiltonian \eqref{eq:H_appendix} gives $\H=\H^{(0)}+\H^{(1)}+\H^{(2)}+\H^{(3)}+\H^{(4)}$, where $\H^{(n)}$ contains all terms in $\H$ proportional to a product of exactly $n$ boson operators.

\subsection{Mean-field/linear spin-wave theory}
\label{supp_section_1A}
The mean-field result is found by considering terms up to $n=2$:

\begin{align}
    \H^{(0)} &= - \frac{1}{2} N S \left(JS (\eps_0 \cos^2\theta + \sin^2\theta \Delta \eps_\Q) + 2 H \sin\theta \right) \\
    \H^{(1)} &= - \frac{J}{2} \sum_{i\not=j} \frac{1}{r_{ij}^3} \left( S\sqrt{2S}\sin\theta\cos\theta (\Delta \eta_i - \eta_j)(\aa_j + \ac_j) \right) - H  \frac{\sqrt{2S}}{2} \cos\theta \sum_i \eta_i (\aa_i + \ac_i) \\
    \H^{(2)} &= - \frac{J}{2} \sum_{i\not=j} \frac{1}{r_{ij}^3} \bigg[ (-2S)(\cos^2\theta + \Delta \eta_i \eta_j \sin^2\theta)\ac_i\aa_i + (\Delta\cos^2\theta + \eta_i \eta_j \sin^2\theta) \frac{S}{2}(\aa_i + \ac_i)(\aa_j + \ac_j) \nn
    &\hspace{.75in} - \frac{S}{2} (\aa_i-\ac_i)(\aa_j-\ac_j) \bigg] + H\sin\theta \sum_i \ac_i \aa_i
\end{align}
where $\eps_\k \equiv \sum_{\r_j \not= \mathbf{0}} \frac{e^{i\k\cdot\r_j}}{|\r_{j}|^3}$ is the dipolar lattice sum for the square lattice, which is an absolutely convergent sum for all $\k$. 
We comment on the methodology used to accurately evaluate these dipolar lattice sums in \ref{dipolar_sum_supp_section}.

By demanding that the ground-state energy $\H^{(0)}$ is minimized, one finds that the classical canting angle $\theta$ is $\sin\theta = H/2JS(\eps_0 - \Delta \eps_\Q)$, which defines the critical field $H_c \equiv 2JS(\eps_0 - \Delta \eps_\Q)$.
Using this expression for $\theta$, all terms in $\H^{(1)}$ vanish, leaving only $\H \approx \H^{(0)} + \H^{(2)}$ to be diagonalized.
We introduce Fourier transformed boson operators
\beq
    \aa_j = \frac{1}{\sqrt{N}} \sum_{\k} e^{i\k\cdot \r_j} \aa_{\k}, \sp \ac_j = \frac{1}{\sqrt{N}} \sum_{\k} e^{-i\k \cdot \r_j} \ac_{\k} \label{eq:Fourier_trans}
\eeq
to write the quadratic Hamiltonian in matrix form as
\begin{align}
    \H &\approx H^{(0)} + H^{(2)} \nn 
    &= E_0 + \frac{JS}{4} \sum_\k \begin{pmatrix}
        \ac_\k & \aa_{-\k}
    \end{pmatrix} \begin{pmatrix}
        A_\k & B_\k \\ B_\k & A_\k
    \end{pmatrix} \begin{pmatrix}
        \aa_{\k} \\ \ac_{-\k}
    \end{pmatrix} \label{eq:H_quad}
\end{align}
where 
\beq
    A_\k = 2\eps_0 - \eps_\k (\Delta \cos^2\theta + 1) - \eps_{\Q+\k} \sin^2\theta, \sp B_\k = - \eps_\k (\Delta \cos^2\theta - 1) - \eps_{\Q+\k} \sin^2\theta. \label{eq:A&B}
\eeq
The Bogoliubov transformation
\beq
    \aa_{\k} = u_\k \ba_\k + v_\k \bc_{-\k}, \sp\ac_{-\k} = v_\k \ba_\k + u_\k \bc_{-\k} \label{eq:Bog_trans}
\eeq
is canonical when $u_\k^2 - v_\k^2 = 1$ and is used to make \eqref{eq:H_quad} diagonal in the $\bc_\k,\ba_\k$ bosons by choosing $u_\k^2 + v_\k^2 = A_\k/\sqrt{A_\k^2 - B_\k^2}$ and $2u_\k v_\k = - B_\k/\sqrt{A_\k^2 - B_\k^2}$
giving the mean-field result
\beq
    \H \approx E_0 + \sum_\k \omega_\k (\bc_\k \ba_\k + 1/2)
\eeq
where
\beq
    \omega_\k = JS \sqrt{(\eps_\k - \eps_0)(\Delta \cos^2\theta \eps_\k + \sin^2\theta \eps_{\k+\Q} - \eps_0)}. \label{eq:stag-FM-disp}
\eeq

At small $|\k|$, $\eps_\k \approx \eps_0 - 2\pi|\k|$ \cite{HPB2012}, and $\eps_{\k+\Q} \approx \eps_\Q + O(|\k|^2)$. Neglecting $O(|\k|^2)$ terms under the radical gives the small-$\k$ expression
\beq
    \omega_\k \approx JS\sqrt{2\pi\big(\eps_0(1-\Delta \cos^2\theta) - \eps_\Q \sin^2\theta\big)|\k|}.
\eeq

\begin{figure}[!h]
    \centering
    \includegraphics[width=1.0\linewidth]{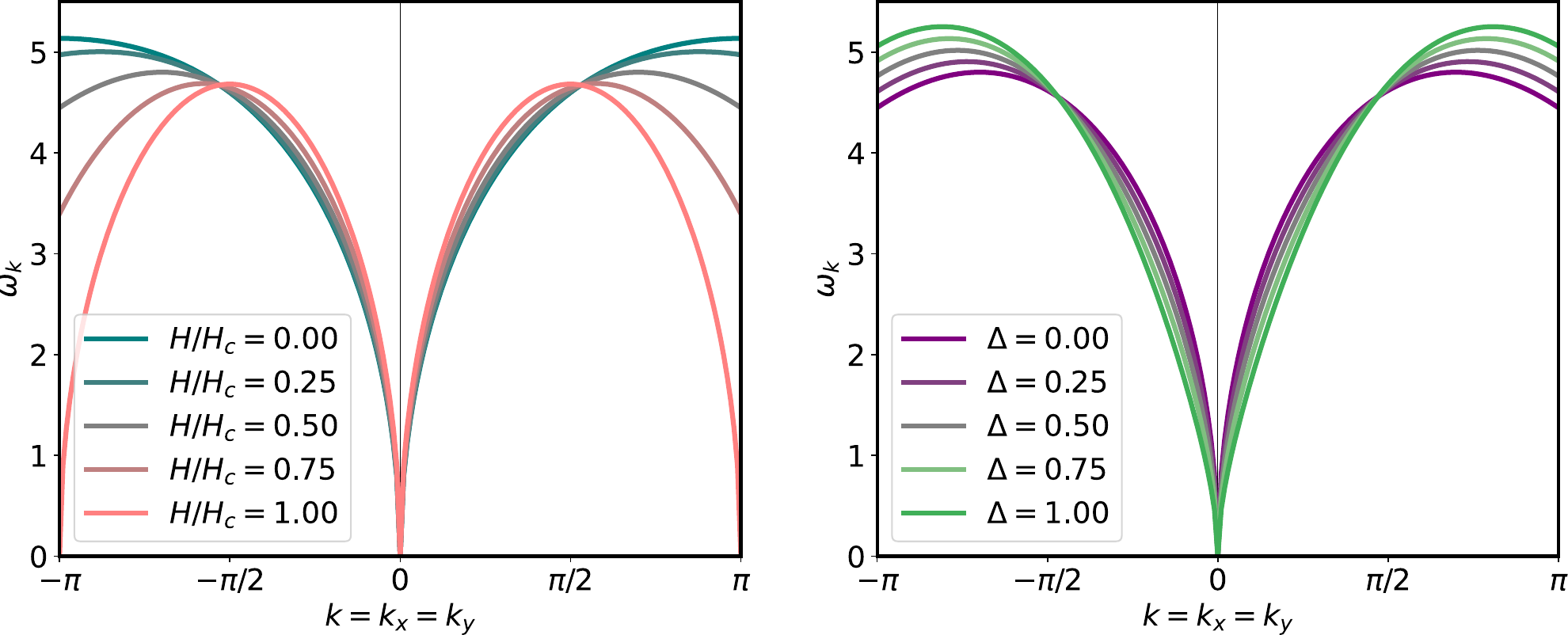}
    \caption{Plots of the staggered field dipolar XXZ FM dispersion. (Left) XY dispersion ($\Delta=0$)  for varying values of the staggered field strength $H/H_c$. (Right) XXZ dispersion at $H/H_c=0.5$ for varying values of the anisotropy $\Delta$. In both panels, the small-$|\k|$ behavior that $\omega_\k \sim \sqrt{|\k|}$ is preserved regardless of $\Delta$ or $H/H_c$.}
    \label{fig:supp_FM_dispersion}
\end{figure}


Similar procedures provide the spin-wave dispersions for the other 3 configurations of the dipolar XXZ Hamiltonian (staggered field AFM, uniform field FM, and uniform field AFM, respectively):
\begin{align}
    \omega_\k^{\text{SF-AFM}} &= JS \sqrt{(\eps_{\k+\Q} - \eps_\Q)(\eps_\k (\Delta \cos^2\theta + \sin^2\theta) -  \eps_\Q ) } \label{eq:SF_AFM_dispersion} \\
    \omega_\k^{\text{UF-FM}} &= JS \sqrt{(\eps_\k - \eps_0)(\eps_\k (\Delta \cos^2\theta + \sin^2\theta) - \eps_0) } \label{eq:UF_FM_dispersion} \\
    \omega_\k^{\text{UF-AFM}} &= JS \sqrt{(\eps_{\k} - \eps_\Q)(\eps_\k \sin^2\theta + \eps_{\k+\Q}\Delta \cos^2\theta - \eps_\Q)} \label{eq:UF_AFM_dispersion}
\end{align}

\subsection{Cubic and quartic interaction terms}
The mean-field result above ignores the 3- and 4-magnon interaction terms:
\begin{align}
    \H^{(3)} &= \frac{J}{2} \sum_{i\not=j} \frac{1}{r_{ij}^3} \sqrt{2S}\sin\theta\cos\theta(\Delta\eta_i - \eta_j)\ac_i\aa_i(\aa_j + \ac_j), \label{eq:H3_realspace} \\
    \H^{(4)} &= -\frac{J}{2} \sum_{i\not=j}\frac{1}{r_{ij}^3} (\cos^2\theta + \Delta \eta_i\eta_j\sin^2\theta) \ac_i\aa_i\ac_j\aa_j . \label{eq:H4_realspace}
\end{align}
To study their effect on the single-magnon spectrum, we must first express them in terms of the Bogoliubov bosons $\bc_\k,\ba_\k$.
Doing so requires one to use the same transformations as above, \eqref{eq:Fourier_trans} and \eqref{eq:Bog_trans}, and to go through some algebra to get \eqref{eq:H3_realspace} into the form of
\beq
    \H^{(3)} = \frac{1}{\sqrt{N}} \sum_{\k,\q} \frac{1}{2!} V_{\k,\q}^{(3)} \bc_\k \ba_\q \ba_{\k-\bar{\q}} + \frac{1}{3!} \bar{V}_{\k,\q}^{(3)} \bc_\k \bc_\q \bc_{-\k-\bar{\q}} + \mbox{H.c.} \label{eq:H3_form}
\eeq
One also gains additonal terms linear in $\bc_\Q,\ba_\Q$ which give a higher-order quantum correction to the canting angle $\theta$ \cite{Chernyshev2009} (which is neglected in our calculations).
In the main text, we restrict ourselves to the XY limit ($\Delta = 0$) of the dipolar XXZ Hamiltonian for FM coupling and in a staggered magnetic field, and we only compute the effect of the cubic interaction terms. 
Thus, we provide the cubic vertex functions for the XY limit of this specific configuration:
\begin{align}
    V^{(3)}_{\k,\q} = \frac{J}{2} \sqrt{2S} \sin\theta \cos\theta \bigg[&\eps_{\k+\Q} (u_\k - v_\k)(u_\q v_{\k-\q+\Q} + v_\q u_{\k-\q+\Q}) \nn
    - &\eps_{\q+\Q} (u_\q - v_\q)(u_\k u_{\k-\q+\Q} + v_\k v_{\k-\q+\Q}) \nn
    - &\eps_{\k-\q} (u_{\k-\q+\Q} - v_{\k-\q+\Q}) (u_\k u_\q + v_\k v_\q) \bigg], \label{eq:V3_decay} \\
    \bar{V}^{(3)}_{\k,\q} = -\frac{J}{2} \sqrt{2S} \sin\theta \cos\theta  \bigg[&\eps_{\k+\Q} (u_\k - v_\k)(u_\q v_{\k-\q+\Q} + v_\q u_{\k-\q+\Q}) \nn 
    + &\eps_{\q+\Q} (u_\q - v_\q)(u_\k v_{\k-\q+\Q} + v_\k u_{\k-\q+\Q}) \nn
    + &\eps_{\k-\q} (u_{\k-\q+\Q} - v_{\k-\q+\Q}) (u_\k v_\q + v_\k u_\q) \bigg] \label{eq:V3_source}
\end{align}
We also give the explicit form of the vertex corresponding to 1-to-3 magnon decays (as defined in Eq. \eqref{eq:Hb} in the main text), showing that none of its terms include an additional momentum $\Q$:
\begin{align}
    V^{(4)}_{\k,\q,\p} = -J \cos^2\theta \bigg[ &\eps_{\k-\q} (u_\k u_\q + v_\k v_\q) (v_\p u_{\k-\q-\p} + u_\p v_{\k-\q-\p}) \nn + &\eps_{\k-\p} (u_\k u_\p + v_\k v_\p) (v_\q u_{\k-\q-\p} + u_\q v_{\k-\q-\p}) \nn  + &\eps_{\q+\p} (u_\k u_{\k-\q-\p} + v_\k v_{\k-\q-\p}) (u_\q v_\p + v_\p u_\q)\bigg] \label{eq:V4}
\end{align}

\subsection{Evaluation of Dipolar lattice sums} 
\label{dipolar_sum_supp_section}
In our calculations, we define
\begin{align}
    \eps_\k \equiv \sum_{\r_j \not= \mathbf{0}} \frac{e^{i \k \cdot \r_j}}{|\r_j|^3}, \label{eq:dipolarsum}
\end{align}
which is an oscillatory sum over the dipolar interactions between a single site and all other sites in the square lattice.
This sum is absolutely convergent in 2 dimensions for any momentum $\k$ (and always real due to lattice symmetries).

Direct computation of $\eps_\k$ as expressed in \eqref{eq:dipolarsum} converges slowly, as terms in the sum decay algebraically with increasing magnitude of lattice vectors.
The sum of slowly converging terms can be transformed into a more rapidly convergent sum using Ewald summation techniques \cite{HPB2012,DelMaestro_Gingras_2004,Enjalran_Gingras_2004} or Epstein zeta functions \cite{EpsteinZeta}.
In all of our numerical calculations working on the square lattice, we compute $\eps_\k$ using the Ewald summation expression given in \cite{HPB2012};
using their expression for $\eps_\k$, we find that sufficient convergence is reached by cutting off the sum beyond 3\textsuperscript{rd}-nearest neighbor lattice vectors.

Working with analogous sums for, \textit{e.g.}, triangular and honeycomb lattices is slightly less straightforward than for the square lattice.
To do so, we make use of the Epstein zeta function as efficiently implemented in \texttt{EpsteinLib} \cite{EpsteinZeta}.
Representing dipolar lattice sums with the Epstein zeta function naturally allows one to introduce a shift to the lattice vectors appearing in the $1/|\r_j|^3$ of \eqref{eq:dipolarsum}, which also makes dealing with strictly inter-sublattice interactions more convenient for our purposes.

\section{Decay threshold boundary for XY FM in staggered fields}

In this section, we provide a formal proof for the two-magnon decay boundary for the dipolar XY FM in a staggered field $H_i = \eta_i H$ on the square lattice. 
Specifically, we show that the kinematic constraint for $1 \to 2$ magnon-decays, 
\begin{align}
    \omega_\k = \omega_\q + \omega_{\k - \q + \Q}, \label{eq:decay_condition_square_lattice}
\end{align}
is satisfied for some $\q$ in the BZ only if $\k$ lies outside the diamond shaped region defined by $|k_x \pm k_y| = \pi$ containing the point $\Gamma$. 

To start, we note that the dispersion $\omega_\k$ is non-negative everywhere and $\omega_{\k} \xrightarrow[]{\k \to 0} 0$. 
Therefore, if $ \omega_\k  > \omega_{\k+\Q}$, by continuity we must have some small $\q$ that satisfies $\omega_\k = \omega_\q + \omega_{\k - \q + \Q}$.
Further, positivity of $\omega_\k$ implies 
$ \omega_\k  > \omega_{\k+\Q} \Leftrightarrow \omega_\k^2 > \omega_{\k+\Q}^2$.
We can substitute the actual dispersion relation to simplify the condition:
\begin{align}
(\omega_\k^2 - \omega_{\k+\Q}^2)/(JS)^2 &= (\eps_0 - \eps_{\k+\Q} \sin^2{\theta})(\eps_0 - \eps_\k) - (\eps_0 - \eps_{\k} \sin^2{\theta})(\eps_0 - \eps_{\k+\Q}) = \eps_0 \cos^2\theta (\eps_{\k+\Q} - \eps_\k) > 0
\end{align}
Since we are considering $H < H_c$, we know that $0 \leq \theta < \pi/2$, meaning that $\eps_0 \cos^2\theta > 0$ ($\eps_0 \approx 9.033$). Therefore, $\omega_\k^2 - \omega_{\k+\Q}^2 > 0$ if and only if $\eps_{\k+\Q} > \eps_\k$.
Thus, the condition for nontrivial two-magnon decays to exist can be simplified into a condition on the dipolar sum $\eps_\k$. 
Rewriting, we have that:
\begin{align} \eps_{\k+\Q} - \eps_{\k} = \sum_{|\r_j|\not=0} = \frac{1}{|\r_j|^3} e^{i\k\cdot\r_j} (e^{i\Q\cdot\r_j} - 1) = \sum_{m,n}^{'} \frac{1}{(m^2 + n^2)^{3/2}} e^{i(k_x m + k_y n)} (e^{i\pi(m+n)}-1) \end{align}
To proceed, we note that 
$e^{i\pi(m+n)}-1=\begin{cases} 0 & 2\mid(m+n) \\ -2 & 2\nmid(m+n) \end{cases}$ \\

Then, we may rewrite the sum again as
\begin{align} \eps_{\k+\Q} - \eps_{\k} = -2 \sum_{\substack{m,n \\ 2\nmid(m+n)}} \frac{1}{(m^2 + n^2)^{3/2}} e^{i(k_x m + k_y n)} \end{align}

From numerical observations, we are inspired to look at the constraint $k_y = \pi - k_x$, giving us:
\begin{align} \eps_{\k+\Q} - \eps_{\k} = -2 \sum_{\substack{m,n \\ 2\nmid(m+n)}} \frac{1}{(m^2 + n^2)^{3/2}} e^{i(k_x (m-n) + \pi n)} \end{align}

At this point we shall consider $m=a$ and $n=b$, which are arbitrary. The term in the sum that these two values for $m,n$ give is 
\begin{align}
\frac{1}{(a^2 + b^2)^{3/2}} e^{ik_x (a-b)}e^{i\pi b}.
\end{align}

To find a set of different values for $m,n$ in the sum whose term cancels this one, consider $m=-b$ and $n=-a$. The term in the sum that these two values for $m,n$ give is 
\begin{align}
  \frac{1}{((-b)^2 + (-a)^2)^{3/2}} e^{i(k_x ((-b)-(-a)) + \pi (-a))} = \frac{1}{(a^2 + b^2)^{3/2}} e^{ik_x (a-b)}e^{-i\pi a}  
\end{align}

By the summation constraint, we know that $2 \nmid (a+b)$, meaning $a$ and $b$ have different parity, which implies that
\begin{align}
    e^{i\pi b} = - e^{i\pi a} = - e^{-i\pi a} 
\end{align}

Therefore, the term in the sum from $m=a, n=b$ is exactly cancelled by the term from $m=-b,n=-a$. Since $a,b$ were arbitrary, this cancellation occurs for all values of $m,n$ where $2\nmid(m+n)$, meaning that $\eps_{\k+\Q} - \eps_{\k} = 0$ under the constraint that $k_y = \pi - k_x$.
Similarly, we can show that the same condition also holds for $k_y = -\pi - k_x$ and $k_y = \pm \pi + \k_x$, thereby completing our proof. 



\subsection{Proof of nearest-neighbor XY decay region boundary}
Following similar logic to that used in the previous subsection, we conclude that the kinematically allowed decays for an arbitrary $\k$ are those where $\q\approx  0$ and $\q\approx\k+\Q$. 
We then propose the condition that if $E_\k > E_{\k+\Q}$, then there exists a nontrivial value of $\q$ for which the two-magnon decay condition is satisfied.

Again, we note that the dispersion $E_\k$ is non-negative everywhere; therefore, $ E_\k > E_{\k+\Q} \Leftrightarrow E_\k^2 > E_{\k+\Q}^2$.
Then there exist nontrivial decays if $E_\k^2 - E_{\k+\Q}^2 > 0$.
We can substitute the actual dispersion to simplify the condition:
\begin{align}
    E_\k^2 - E_{\k+\Q}^2 &= (\gamma_0 + \gamma_{\k} \sin^2{\theta})(\gamma_0 - \gamma_\k) - (\gamma_0 + \gamma_{\k+\Q} \sin^2{\theta})(\gamma_0 - \gamma_{\k+\Q}) \nn
    &= \gamma_0 (1 - \sin^2\theta)(\gamma_{\k+\Q} - \gamma_\k) + \sin^2\theta (\gamma_{\k+\Q}^2 - \gamma_{\k}^2) \tag{note that $\gamma_{\k+\Q}^2 = \gamma_{\k}^2$} \nn
    &= \gamma_0 (1 - \sin^2\theta)(\gamma_{\k+\Q} - \gamma_\k) > 0
\end{align}

Since we are considering $\delta < \delta_c$, we know that $0 \leq \theta < \pi/2$, meaning that $1 - \sin^2\theta > 0$. Also, $\gamma_0 = 4$, so $\gamma_0 (1 - \sin^2\theta) > 0$. So $E_\k^2 - E_{\k+\Q}^2 > 0$ if and only if $\gamma_{\k+\Q} > \gamma_\k$.

Thus, the condition for nontrivial two-magnon decays to exist can be simplified into a condition on the nearest-neighbor sum $\gamma_\k$. The curve along which $\gamma_\k = \gamma_{\k+\Q}$ is the curve along which the decay region boundary lies, since $\gamma_\k$ monotonically decreases, and $\gamma_{\k+\Q}$ monotonically increases, from $0$ to $\pi$. Again, we look at the line $k_y = \pi - k_x$ and observe that
\begin{align}
    \gamma_\k &= 2\big[\cos{k_x} + \cos{(\pi - k_x)} \big] \nn
    \gamma_{\k+\Q} &= 2\big[\cos{(k_x + \pi)} + \cos{(\pi - k_x + \pi)}\big] = 2\big[\cos{(k_x - \pi)} + \cos{k_x}\big] = \gamma_\k
\end{align}
Thus, we see that the decay region boundary is along the line $k_y = \pi - k_x$.

\section{Regularization of decay singularities}
In this section, we provide additional details discussing the origin and regularization of singularities in the magnon self-energy, and consequently the decay rate. 

\subsection{Origin of decay singularities}
\begin{figure}[!h]
    \centering
    \includegraphics[width=1.0\linewidth]{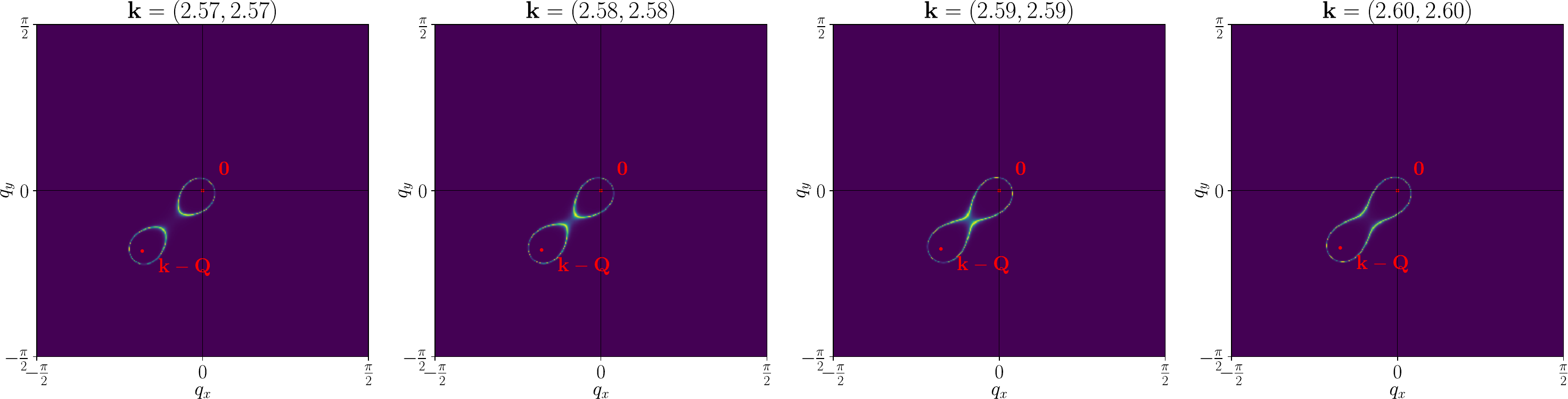}
    \caption{Plots of decay contours in $\q$-space for the staggered-field XY FM at $H/H_c = 0.5$ for four values of $\k$ along the BZ diagonal (increasing from left to right). A topological transition in the decay contours corresponds to crossing the Van Hove singularity (ring-like feature in Fig. \ref{fig3:stag_XY_FM}(a)), which occurs at $\k$ in between $(2.57,2.57)$ and $(2.59,2.59)$.}
    \label{fig:decay_contours_FM}
\end{figure}

Van Hove singularities in the 2-magnon DoS are inherited by the one loop on-shell magnon self-energy $\Sigma(\k,\omega_\k)$.
In our scenario, such singularities are caused by the single-magnon dispersion $\omega_\k$ intersecting the two-particle continuum $E_2(\k,\q)$ at either (i) a minimum or (ii) a saddle-point in $\q$-space.
In the staggered-field dipolar XY FM, only the latter kind of Van Hove singularity is present.
Figure \ref{fig:decay_contours_FM} shows a sequence of decay contours in $\q$-space as a function of the initial magnon momentum $\k$, for the staggered-field XY FM on the square lattice.
As $\k$ increases along the diagonal, the two ring-like decay contours seen in Figure \ref{fig3:stag_XY_FM}(a) merge, and the set of $\q$ for which 2-magnon decays are allowed transitions from a single contour to two disconnected contours enclosing $\q=\mathbf{0}$ and $\q=\k-\Q$.

\subsection{Regularization of decay singularities via off-shell Dyson's equation}
In the case of the dipolar XY FM on the square lattice, although an initial magnon with momentum $\k$ in the decay region is unstable, the momenta of two outgoing magnons into which it may decay are guaranteed to lie outside of the decay region; note that the decay contours in $\q$-space seen in Fig. \ref{fig:decay_contours_FM} lie far outside the decay region seen in Fig. \ref{fig2:decay_regions}(a). 
Therefore, the two outgoing magnons of a 1-to-2 magnon decay acquire no imaginary part in the one loop approximation.
Thus, the conditions of this logarithmic singularity are similar to those studied in Ref. \cite{Chernyshev2009}, so we regularize the singularities in our model by solving Dyson's equation off-shell in the same manner.
Then, by accounting for the finite lifetime of the initial magnon, we numerically solve the equations \eqref{eq:offShellDyson} self-consistently.
In practice, we do so by first setting $\bar{\omega}_\k=\text{Re}[\Sigma(\k,\omega_\k + i\epsilon)]$ and $\Gamma_\k=-\text{Im}[\Sigma(\k,\omega_\k + i\epsilon)]$ where $\epsilon$ represents a small Lorentzian broadening. 
Then we iterate \eqref{eq:offShellDyson} until the average difference between the current and previous solutions is less than one part in $10^5$.

\begin{figure}[!h]
    \centering
    \includegraphics[width=0.5\linewidth]{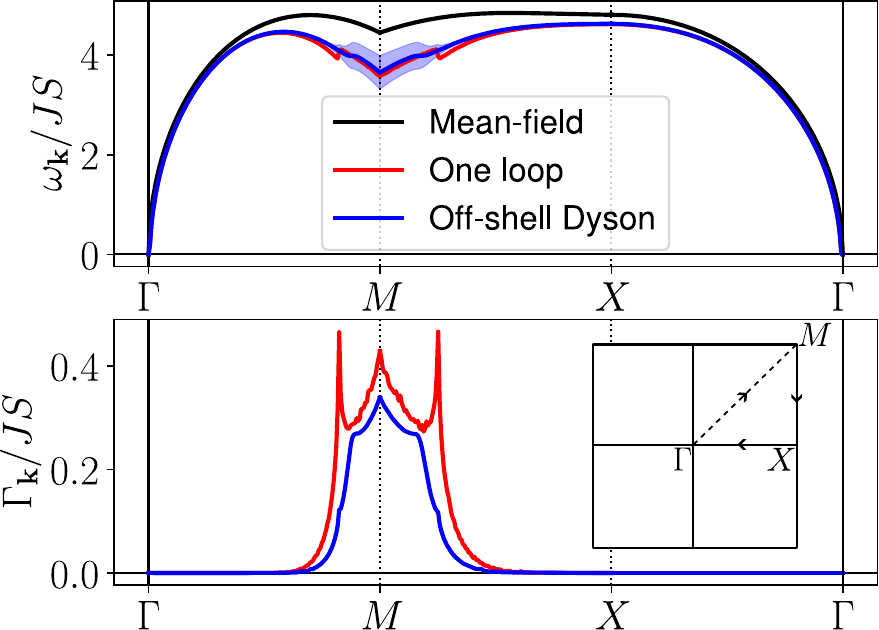}
    \caption{Plots of magnon dispersions (top) and decay rates (bottom) over a 1D path through the BZ (bottom inset).}
    \label{fig:lineCuts_supp}
\end{figure}

The result of doing so, using an $800\times800$ $\q$-grid for integration but evaluating over an $81\times81$ $\k$-grid and setting $\epsilon= 10^{-2} JS$, is given in Fig. \ref{fig3:stag_XY_FM}(b). 
Insets of the figure were obtained using a $1000\times1000$ $\q$-grid, setting $\epsilon=10^{-3} JS$, and plotted over just the $\Gamma M$ line cut in the BZ, but plots with the same parameters over an extended path is given in Fig. \ref{fig:lineCuts_supp}. 

\section{Threshold behavior}
In this section, we provide a detailed derivation of the magnon-decay rate at the threshold, i.e., near the decay boundary for the dipolar XY FM in a staggered field.
Because of the $C_4$ symmetry of the problem, we can restrict ourselves to the first quadrant of the BZ. 
Consider a magnon with initial momentum $\k = \k_0 + \p$, where $\k_0$ is a point on the decay boundary (i.e., $k_{0,x} + k_{0,y} = \pi$), and $|\p|$ is small. 
In the one-loop approximation, its decay rate is given by
\begin{equation}
\Gamma_{\k = \k_0 + \p} = \pi \int \frac{d^2 q}{(2\pi)^2} \, |V^{(3)}_{\k_0 +\p ,\q}|^2 \, \delta(\omega_{\k_0 + \p} - \omega_\q - \omega_{\k_0 + \p -\q+\Q})
\end{equation}
It is intuitively expected and numerically confirmed that the magnons near the threshold decay by emitting two magnons, one of which is a Goldstone magnon and the other one is a near-threshold magnon. 
W.l.o.g., we can hence evaluate the integral when $\q$ is small, and multiply the final result by 2. 
Taylor expanding the argument of the delta function (that imposes energy conservation) everywhere except $\q = 0$ where it is non-analytic, we find that 
\begin{equation}
\omega_{\k_0 + \p} - \omega_\q - \omega_{\k_0 + \p -\q+\Q} \approx (\v_{\k_0} + \v_{\k_0 + \Q})\cdot \p - \v_{\k_0 + \Q} \cdot \q - \bar{c} \sqrt{|\q|} = (v_x + v_y)(p_x + p_y) - (v_y q_x + v_x q_y) - \bar{c} \sqrt{|\q|} 
\end{equation}
where $\bar{c} = JS \sqrt{2 \pi (\varepsilon_0 - \varepsilon_\Q \sin^2\theta )}$, and in the last step we defined $v_{\k_0} \equiv (v_x, v_y)$, then $\v_{\k_0 + \Q} = - (v_y, v_x)$ using $\v_\k$ is an odd function of $\k$ and its reflection about the diagonal $k_x = k_y$ interchanges the x and y components.
We note that the argument of the delta function only depends on $p_x + p_y = \sqrt{2} \p \cdot \hat{\Q}$. i.e., the normal distance of the point $\k$ from the decay threshold line.
Further, for small $|\q|$, the integrand is dominated by the square root dispersion and the linear term in $q_x, q_y$ can be neglected, leading to the following simplification for the decay rate:
\begin{equation}
\Gamma_{\k = \k_0 + \p} \approx \frac{2 \pi}{\sqrt{2}(v_x + v_y)} \int d^2\q \, |V^{(3)}_{\k_0 +\p ,\q}|^2 \, \delta( \p \cdot \hat{\Q} - c\sqrt{|\q|})
\end{equation}
which is Eq.~\eqref{eq:threshold} in the main text.

From the above analysis, it is evident that we only need to consider the cubic vertex squared for momenta $\q$ that satisfy $c\sqrt{q} = \p \cdot \hat{\Q}$, which we turn to next.
We begin with the definition of the vertex. 
\[ V^{(3)}_{\k,\q} = C \left[ f(\k,\q) - g(\k,\q) - g(\k,\k-\q+\Q) \right] \]
where
\begin{align*}
    f(\k,\q) &\equiv \eps_{\k+\Q} (u_\k - v_\k)(u_\q v_{\k-\q+\Q} + v_\q u_{\k-\q+\Q}) \\
    g(\k,\q) &\equiv \eps_{\q+\Q} (u_\q - v_\q)(u_\k u_{\k-\q+\Q} + v_\k v_{\k-\q+\Q}) \\
    g(\k,\k-\q+\Q) &= \eps_{\k-\q} (u_{\k-\q+\Q} - v_{\k-\q+\Q}) (u_\k u_\q + v_\k v_\q)
\end{align*}
and $C = \frac{J}{2} \sqrt{2S} \sin\theta \cos\theta $ is a complex constant.
Now, we consider a point near the decay threshold $\k = \k_0 + \p$, where $\k_0$ lies on the decay boundary and $|\p|$ is small.
Since $|\p|$ is small, we also need $|\q|$ (or $|\k - \q + \Q|$) to be small for energy conservation to be satisfied. 
Since the vertex is symmetric under $\q \to \k - \q + \Q$, we can only look at the case where $|\q|$ is small.
In this limit, we expand each of the functions above, and keep only the leading order term in $|\q|$.
If the term can be Taylor expanded, we will only keep the first (constant) term, while if it is non-analytic we will keep the leading non-analytic dependence. 
We will come back to examine the validity of this later - it will turn out that is not quite valid (we need to keep derivative terms) but it still gives us the correct scaling. 

We first note that,
\beq
u_\q - v_\q = \left( \frac{\varepsilon_0 - \varepsilon_\q}{\varepsilon_0 - \varepsilon_{\q + \Q} \sin^2\theta }\right)^{1/4} \implies u_{\k_0} - v_{\k_0} = u_{\k_0 + \Q} - v_{\k_0 + \Q}
\eeq
since $\varepsilon_{\k_0 + \Q} = \varepsilon_{\k_0}$ for $\k_0$ lying on the decay boundary. 
This, together with the constraint $u_\q^2 - v_\q^2 = 1$ for all $\q$ implies that $u_{\k_0} + v_{\k_0} = u_{\k_0 + \Q} + v_{\k_0 + \Q}$.
Combining these two results, we find that $u_{\k_0} = u_{\k_0 + \Q}$ and $v_{\k_0} = v_{\k_0 + \Q}$ when $\k_0$ lies on the boundary. 
We also note that for small $q$,
\beq
u_\q - v_\q \approx \left( \frac{\varepsilon_0 - \varepsilon_\q}{\varepsilon_0 - \varepsilon_{\Q} \sin^2\theta }\right)^{1/4} \approx \left( \frac{2 \pi a |\q|}{\varepsilon_0 - \varepsilon_{\Q} \sin^2\theta }\right)^{1/4} \implies u_\q + v_\q  = \frac{1}{u_\q - v_\q } \approx \left( \frac{\varepsilon_0 - \varepsilon_{\Q} \sin^2\theta }{2 \pi a |\q|}\right)^{1/4}
\eeq
Thus, both $u_\q$ and $v_\q$ will diverge as $|\q|^{-1/4}$ for small $|\q|$, but the combination $u_\q - v_\q \sim |\q|^{1/4}$ does not diverge.

Using the simplifications above, we have, to leading order
\begin{align*}
    f(\k_0 + \p,\q) &=  \eps_{\k_0+\p + \Q} (u_{\k_0 + \p} - v_{\k_0 + \p})(u_\q v_{\k_0 + \Q + \p -\q} + v_\q u_{\k_0 +\Q + \p - \q}) \approx  \eps_{\k_0} (u_{\k_0} - v_{\k_0})(u_\q v_{\k_0} + v_\q u_{\k_0}) \\
    g(\k_0 + \p,\q) &= \eps_{\q+\Q} (u_\q - v_\q)(u_{\k_0 + \p} u_{\k_0 + \p -\q+\Q} + v_{\k_0 + \p} v_{\k_0 + \p -\q+\Q}) \approx \varepsilon_{\Q}(u_\q - v_\q)(u^2_{\k_0} + v^2_{\k_0} )  \\
    g(\k_0 + \p,\k_0 + \p-\q+\Q) &= \eps_{\k_0 + \p -\q} (u_{\k_0 + \p -\q+\Q} - v_{\k_0 + \p -\q+\Q}) (u_{\k_0 + \p} u_\q + v_{\k_0 + \p} v_\q) \approx \eps_{\k_0} (u_{\k_0} - v_{\k_0}) (u_{\k_0} u_\q + v_{\k_0} v_\q)
\end{align*}
In all of the above, we have expanded in a Taylor series in $\q$ and $\p - \q$, which are both small, except the terms are non-analytic at small $\q$, and also repeatedly used $\eps_{\k_0} = \eps_{\k_0 + \Q}, u_{\k_0} = u_{\k_0 + \Q}, v_{\k_0} = v_{\k_0 + \Q}$. 
Within this approximation:
\beq
 f(\k_0 + \p,\q)  -  g(\k_0 + \p,\k_0 + \p-\q+\Q) \approx - \eps_{\k_0} (u_{\k_0} - v_{\k_0})^2 (u_\q - v_\q)
\eeq
Therefore, the vertex is given by
\beq
V^{(3)}(\k_0 + \p,\q) \approx  - C \left[ \eps_{\k_0} (u_{\k_0} - v_{\k_0})^2 + \eps_\Q (u^2_{\k_0} + v^2_{\k_0} ) \right] (u_\q - v_\q) 
\eeq
In this approximation, therefore, $|V^{(3)}(\k_0 + \p,\q)|^2$ scales as $ (u_\q - v_\q)^2 \approx q^{1/2}$. 

Next, we consider the terms which we neglected in the Taylor expansion, and show that we cannot neglect them altogether.
The reason is as follows: the energy conservation constraint sets $\sqrt{q} \sim p$, implying that a linear term in $p$ in the Taylor expansion may multiply $u_\q \sim q^{-1/4}$ or $v_\q \sim q^{-1/4}$ to give a term $\sim q^{1/4}$ in $V^{(3)}(\k_0 + \p, \q)$.
Keeping leading order terms in $\p$ in the Taylor expansion (but not in $\q$, as these would give subleading corrections) and carrying out the tedious algebra, we find the linear in $p$ term to be
\beq
\p \cdot(\v_{\k_0 + \Q}- \v_{\k_0})\bigg[ (u_{\k_0} - v_{\k_0})(u_\q u_{\k_0} + v_\q v_{\k_0}) + \frac{(u_{\k_0} - v_{\k_0})[\eps_0(1 + \sin^2\theta) - 2 \eps_{\k_0} \sin^2\theta]}{4(\eps_0 - \eps_{\k_0})(\eps_0 - \eps_{\k_0} \sin^2\theta)} \eps_{\k_0}(u_\q u_{\k_0} + v_\q v_{\k_0}) \nn 
+ \eps_{\k_0} (u_\q + v_\q)\frac{[\eps_0(1 + \sin^2\theta) - 2 \eps_{\k_0} \sin^2\theta]}{8(\eps_0 - \eps_{\k_0})(\eps_0 - \eps_{\k_0} \sin^2\theta)} \bigg] \nn
\eeq
Since $u_\q$ and $v_\q$ scale as $q^{-1/4}$, for $p \sim \sqrt{q}$ from the kinematic constraint, both the p-independent and p-linear term in the vertex $V^{(3)}(\k_0 + \p,\q)$ scale as $q^{1/4}$.
Thus, we have derived that $|V^{(3)}(\k_0 + \p,\q)|^2$ asymptotically scales as $ \sqrt{q}$ near the threshold for $p \sim \sqrt{q}$, used in the main text. 
We have also shown that the singular part of the term linear in $\p$ is proportional to $\p \cdot (\v_{\k_0} - \v_{\k_0 + \Q})$, implying that at this level of approximation the vertex depends only on the component of $\p$ normal to the decay boundary, as claimed in the main text. 

\section{Dynamical structure factor}
In this section, we provide detailed expressions for the dynamical spin-structure factor in terms of the renormalized magnon Green's functions $G(\q,\omega)$. 
We note that our expressions for twisted-frame spin correlation functions are similar to Ref.~\onlinecite{Mourigal2010}, the main differences being that our ordering direction is along the x-axis and that we neglect Hartree-Fock correction factors:
\begin{align}
    S^{xx}(\k,\omega) &= -\frac{1}{N} \sum_\q (u_{\k-\q} v_\q + u_\q v_{\k-\q})^2 \text{Im} \left\{ \int_{-\infty}^{\infty} dx G(\q,x) G(\k-\q,\omega-x)\right\}, \label{eq:Sxx} \\
    S^{yy}(\k,\omega) &= -S(u_\k + v_\k)^2 \text{Im}G(\k,\omega), \label{eq:Syy} \\
    S^{zz}(\k,\omega) &= -S(u_\k - v_\k)^2 \text{Im}G(\k,\omega). \label{eq:Szz}
\end{align}
The correlation functions above are for twisted-frame spin operators, so we move back to the lab frame using \eqref{eq:twisted_frame}:
\begin{align}
    S^{x_0 x_0} (\k,\omega) &= S^{xx}(\k,\omega) \cos^2\theta + S^{zz}(\k+\Q,\omega)\sin^2\theta,\nn
    S^{y_0 y_0} (\k,\omega) &= S^{yy}(\k,\omega), \nn
    S^{z_0 z_0} (\k,\omega) &= S^{xx}(\k+\Q,\omega) \sin^2\theta + S^{zz}(\k,\omega)\cos^2\theta.
\end{align}

\def\wbar{\bar{\omega}}

Evaluating the integral in $S^{xx}(\k,\omega)$ is generally challenging, but the form of $G(\k,\omega)$ is simplified by the fact that we get a frequency-independent self-energy from solving Dyson's equation off-shell.
Letting $\Sigma(\k,\omega) = \Sigma(\k) = \wbar_\k - i\Gamma_\k$,
\begin{align} 
G(\q,x) = \frac{1}{x - \wbar_\q + i\Gamma_\q}, \sp 
G(\k-\q,\omega - x) = \frac{1}{\omega - x - \wbar_{\k-\q} + i\Gamma_{\k-\q}}.
\end{align}
Then defining $\tau = x - \wbar_\q$, the integrand in \eqref{eq:Sxx} becomes
\begin{align}
    \frac{1}{\tau + i\Gamma_\q} \frac{1}{(\omega - \wbar_\q - \wbar_{\k-\q} - \tau) + i\Gamma_{\k-\q}} = \frac{1}{\tau + i\Gamma_\q} \frac{1}{(\omega' - \tau) + i\Gamma_{\k-\q}}
\end{align}
where $\omega' \equiv \omega - \wbar_\q - \wbar_{\k-\q}$. 
Then,
\begin{align}
    \text{Im} \left\{ \int_{-\infty}^{\infty} dx G(\q,x) G(\k-\q,\omega-x)\right\} &= \int_{-\infty}^{\infty} d\tau \left[ \Gamma_\q \omega' - (\Gamma_\q - \Gamma_{\k-\q})\tau \right] \frac{1}{(\tau^2 + \Gamma_\q^2)} \frac{1}{((\omega'-\tau)^2 + \Gamma_{\k-\q}^2)} \nn
    &= \frac{-\pi \omega'}{\Gamma_{\k-\q}} \left\{ \frac{\Gamma_\q + \Gamma_{\k-\q}}{\omega'^2 + (\Gamma_\q + \Gamma_{\k-\q})^2} - \frac{\Gamma_\q - \Gamma_{\k-\q}}{\omega'^2 + (\Gamma_\q - \Gamma_{\k-\q})^2} \right\}. \label{eq:final_convolution}
\end{align}
The integral in $S^{xx}(\k,\omega)$ becomes the difference of two integrals: a convolution of two Lorentzians and the same thing with one power of $\tau$ in the integrand.
Both of these integrals are analytically evaluable, giving us the final expression in \eqref{eq:final_convolution}.
Using it, we compute $S(\k,\omega)$ as defined in the main text using a $600\times600$ $\q$-grid for 50 values of $\k$ along the $\Gamma M$ line cut, evaluating at 100 values of $\omega$ for each $\k$.
The result is shown as a log plot in Fig. \ref{fig1:schematic}(c) in the main text.

\section{Quartic terms and Three-magnon decays}
\label{supp_section_quartic_terms_3magnon_decays}
\subsection{Contribution of quartic terms to one-loop calculations}
In our discussion of Eq. \eqref{eq:Hb} in the main text, we note that the contributions of 4-magnon decay processes are higher order in $1/S$ than the 3-magnon contributions, and thus we neglect them in our calculations.
However, the quartic terms do contribute order-$1/S$ corrections to the mean field dispersion $\omega_\k$, which are obtained via a Hartree-Fock (HF) decoupling procedure \cite{ZC_RMP2013}.
Such a procedure corrects only the real part of the dispersion, giving $\tilde{\omega}_\k = \omega_\k + E_\k^{(4)}$, where
\begin{align}
    E_\k^{(4)} = \left( u_\k^2 + v_\k^2 \right) \cdot \delta A_\k - 2 u_\k v_\k \cdot \delta B_\k. \label{eq:E_k^4}
\end{align}
For the dipolar XY FM in a staggered field on a square lattice (the configuration of interest in Fig. \ref{fig3:stag_XY_FM}), the terms $\delta A_\k$ and $\delta B_\k$ take the form
\begin{align}
    \delta A_\k = \frac{1}{2} \sum_\q (\eps_0 + \eps_{\k+\q}) \left( \frac{A_\q}{\omega_\q} - 1 \right), \hspace{10pt} \delta B_\k = \sum_\q \eps_{\k + \q} u_\q v_\q.\label{eq:deltaA_deltaB_XY_FM}
\end{align}

\begin{figure}[!h]
    \centering
    \includegraphics[width=0.5\linewidth]{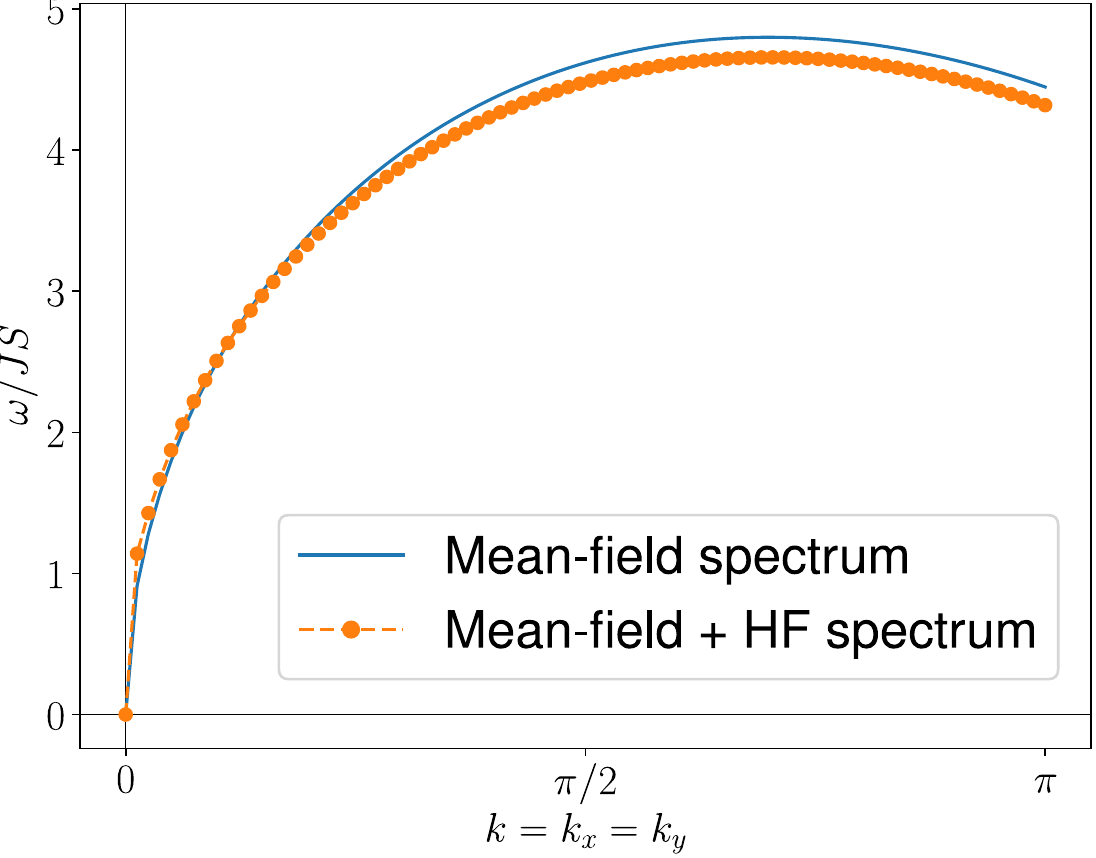}
    \caption{Plot showing mean-field dispersion $\omega_\k$ (blue) and Hartree-Fock corrected dispersion $\tilde{\omega}_\k$ (orange) for the dipolar XY FM in a staggered field on the square lattice, both taken along the $\Gamma M$ line cut.}
    \label{fig:HF_corrections}
\end{figure}

Figure \ref{fig:HF_corrections} shows that the HF corrections to the mean-field dispersion, given explicitly in Eqs. \eqref{eq:E_k^4} and \eqref{eq:deltaA_deltaB_XY_FM}, do not qualitatively nor quantitatively affect the dispersion at $\k$ far from the $\Gamma$ point (here, we find the corrections lower $\omega_\k$ by less than 3 percent).
We find that the HF corrections are more significant for $\k$ near $\Gamma$, but they do not alter the characteristic behavior $\omega_\k \sim \sqrt{|\k|}$.
Therefore, because we are primarily concerned with the effect of cubic anharmonicities on magnon stability, we neglect HF corrections in the one-loop calculations in the main text.

\subsection{Three-magnon decays in dipolar AFMs}
In the main text, we predict that antiferromagnetic magnons may be stabilized in the Heisenberg limit, but the one-loop decay rates shown in Figure \ref{fig4:AFM_decays} only take into account contributions from 1 to 2 magnon decays.
Then, to justify this prediction, we briefly examine the on-shell 3-magnon density of states (DoS) $\rho_3(\k)$, defined as
\begin{align}
    \rho_3(\k) \equiv \frac{1}{N^2} \sum_{\q,\p} \delta(\omega_\k - \omega_{\q} - \omega_{\p} - \omega_{\k-\q-\p}).
\end{align} 
This quantity is closely related to contributions from the quartic vertex \eqref{eq:V4} to the magnon decay rate $\Gamma_\k$ at zero temperature. 
The leading order two-loop contribution is as follows \cite{ZC_RMP2013}:
\begin{align}
    \Gamma_\k^{(4)} \sim \sum_{\q,\p} |V_{\k,\q,\p}^{(4)}|^2 \delta(\omega_\k - \omega_{\q} - \omega_{\p} - \omega_{\k-\q-\p}).
\end{align}

\begin{figure}[!h]
    \centering
    \includegraphics[width=1.0\linewidth]{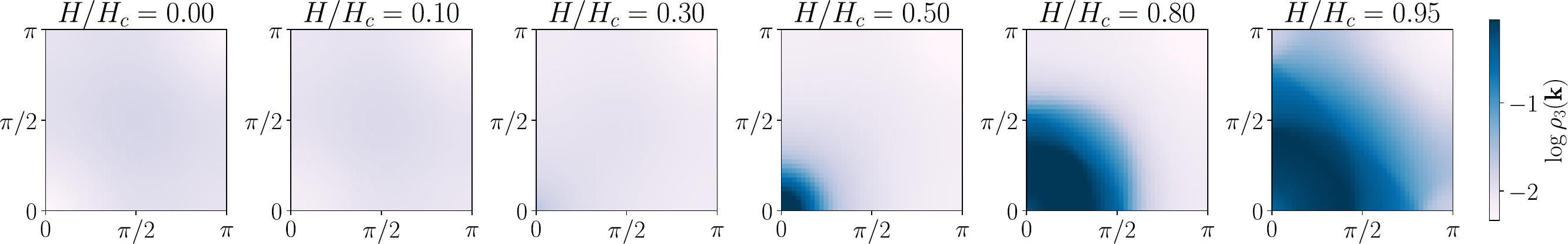}
    \caption{Plots of $\log[\rho_3(\k)]$ for the dipolar HAFM in a uniform field for multiple values of $H/H_c$ (increasing from left to right).}
    \label{fig:3magnon_DOS_uniform_HAFM}
\end{figure}

Note that the uniform-field XXZ AFM dispersion \eqref{eq:UF_AFM_dispersion} is gapless at $\Q$. 
Since the vertex $V_{\k,\p,\q}^{(4)}$ is not expected to be singular at $\k 
= \Q$, the stability of low-energy magnons requires a small DoS at $\k=\Q$.
Figure \ref{fig:3magnon_DOS_uniform_HAFM} shows an increased DoS near $\k=\mathbf{0}$ with increased $H/H_c$, but it remains suppressed for the low-energy magnons at $\Q$.
This indicates that the low-energy magnons of the HAFM in a uniform field are stable against 1 to 3 magnon decays.
Taken together with the fact that 1 to 2 magnon decays are kinematically forbidden at $H/H_c < 0.29$ (Table.~\ref{tab}), we conclude that the low-energy magnons of the HAFM in a uniform field are perturbatively stable at small $H$, as claimed in the main text. 

\section{Effect of different lattice geometries}

Throughout this work, we have limited our discussion to the dipolar XXZ model in an external magnetic field with sites arranged on a \textit{square lattice}. 
One may wonder the extent to which our results carry over to other 2D lattice geometries.

The choice of lattice geometry directly affects the ground state of the dipolar XXZ Hamiltonian \eqref{eq:H_appendix}, which (a) may not have collinear order even at $H=0$ and (b) may not be well-approximated by the classical ground state (which is used in our spin-wave analysis).
Furthermore, to have a straightforward and analogous way of implementing a staggered magnetic field as we did on the square lattice, we can only consider bipartite lattices for staggered fields, while we may consider both bipartite and non-bipartite lattices for uniform fields.
\vspace{0.5cm}

\subsection{Uniform-field dipolar XXZ FM on the triangular/honeycomb lattices}

For a dipolar XXZ FM in a \textit{uniform} field, neither of the two caveats mentioned above appears to raise an issue.
The long-range FM coupling ensures collinear order in the $x_0y_0$-plane at zero field, and turning on a finite field along $\pm \hat{z}_0$ cants the spin at each site out of plane in the same direction.
For example, we find numerically that kinematic constraints prevent nontrivial magnon decays at any field $0 \leq H \leq H_c$ in the uniform-field dipolar XY FM on the triangular lattice (similarly to the same model on the square lattice; see the main text for discussion). 
Thus, magnons are stable in two-dimensional dipolar XY ferromagnets, independent of the specific lattice. 

\subsection{Staggered-field dipolar XY FM on the honeycomb lattice}
Here, we briefly mention some results on the staggered-field dipolar XY FM on the honeycomb lattice, which is bipartite.
The classical ground state is taken to be the appropriate analog of Figure \ref{fig1:schematic}(a), where spins on each sublattice cant in the direction of the sublattice-dependent applied field.

To use spin-wave theory, we define two different Holstein-Primakoff boson species $c_i,c_i^\dag$ and $d_i,d_i^\dag$ which live on sublattice $1$ and $2$, respectively, of the honeycomb lattice.
We then follow the procedure outlined in Section \ref{supp_section_I} by Fourier transforming and diagonalizing the Hamiltonian via a canonical Bogoliubov transformation, yielding a mean-field magnon dispersion with two inequivalent branches $\omega_\k^+$ and $\omega_\k^-$, both defined in the 1st BZ.
The lower branch has the small-$\k$ behavior $\omega_\k^- \sim \sqrt{|\k|}$, and $\omega_\k^- \leq \omega_\k^+$ for all $\k$. 
The only nontrivially satisfiable decay condition is then 
\begin{align}
    \omega_\k^+ &= \omega_\q^- + \omega_{\k-\q}^- .\label{eq:decay_condition_honeycomb}
\end{align}

\begin{figure}[!h]
    \centering
    \includegraphics[width=1.0\linewidth]{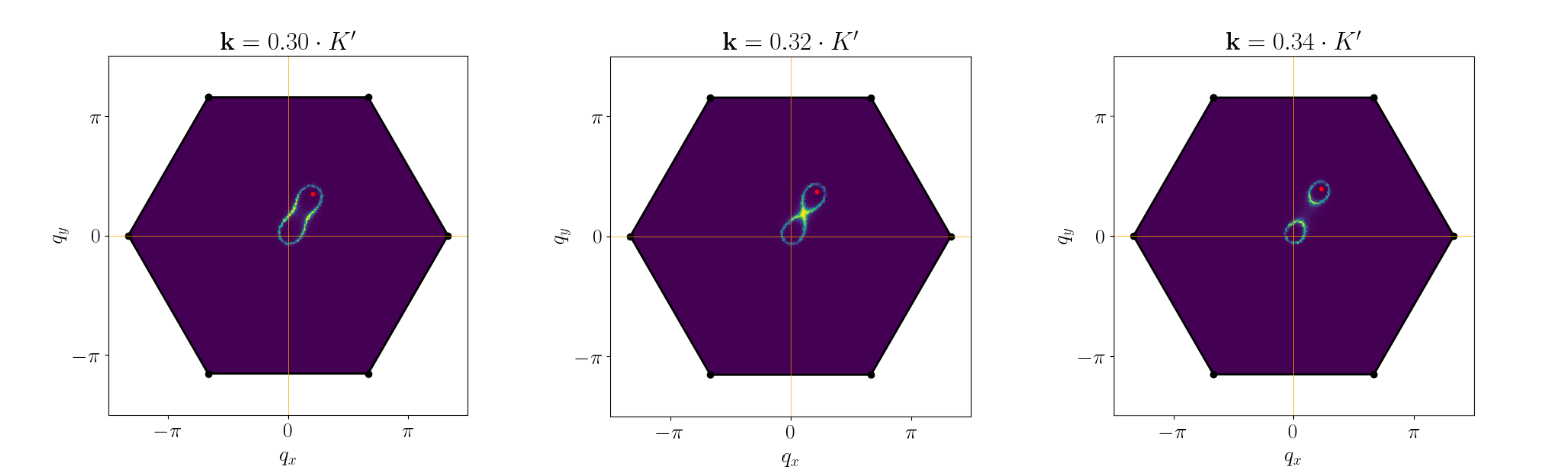}
    \caption{Plots of decay contours in $\q$-space for the staggered-field XY FM on the honeycomb lattice at $H/H_c = 0.5$ for three values of $\k$ along the $\Gamma K'$ line cut.
    Red dots indicate the momentum $\k$ of the incoming magnon.
    Dipolar lattice sums were computed using Epstein zeta functions (see Section \ref{dipolar_sum_supp_section}).}
    \label{fig:honeycomb_decay_plots}
\end{figure}

Figure \ref{fig:honeycomb_decay_plots} shows the evolution of the two-magnon decay contours as the momentum $\k$ of an initial magnon increases along the $\Gamma K'$ line cut.
Note the similarity between these plots and those for the same model on a square lattice shown in Figure \ref{fig:decay_contours_FM}. 
The only obvious qualitative difference is that in the latter case the contours encircle momenta $\q=\mathbf{0}$ and $\q=\k - \Q$, but here they encircle momenta $\q=\mathbf{0}$ and $\q=\k$.
This can be attributed to the fact that, when working on the honeycomb lattice, we traded the decay condition with a momentum shift \eqref{eq:decay_condition_square_lattice} for a decay condition \eqref{eq:decay_condition_honeycomb} involving the two different branches of the dispersion.

\subsection{Staggered-field dipolar XY AFM on the honeycomb lattice}
We also briefly comment on some qualitative results for the staggered-field dipolar XY AFM on the honeycomb lattice.
Following essentially the same procedure used for the FM in the previous subsection, we obtain two inequivalent branches of the magnon dispersion, $\omega_\k^+$ and $\omega_\k^-$.
The lower branch is gapless at $\k=0$, and the leading-order behavior for small $\k$ is $\omega_\k^- \sim |\k|$ for all strengths of the staggered field.

Consequently, at arbitrary staggered field strength $H$, there are two possible decay conditions which can both have nontrivial solutions: 
(i) $\omega_\k^- = \omega_\q^- + \omega_{\k-\q}^-$ and (ii) $\omega_\k^+ = \omega_\q^- + \omega_{\k-\q}^-$.
We observe numerically that the region of momenta in the BZ for which these conditions are satisfied varies as a function of $H$, and furthermore the regions for the two conditions evolve differently with $H$. \\

\textit{Condition (i).} 
At staggered field strengths $H \ll H_c$, $\omega_\k^- \sim |\k|$ everywhere in the first BZ, so the decay contours for condition (i) appear as straight lines in $\q$-space connecting the $\Gamma$-point to the incoming magnon momentum $\k$.
However, at larger fields, the linearity of the dispersion holds only at smaller values of $\k$, so condition (i) cannot be satisfied nontrivially. 
\\

\textit{Condition (ii).}
This condition is satisfied for a large but slightly varying region of momenta $\k$ at all $H \leq H_c$ (and thus there is no threshold field for decays).
The decay contours undergo topological transitions, due to van Hove singularities in the 2-magnon DoS, in a manner qualitatively similar to magnon decays in the FM.